%% file: suprot_revised_vs2.tex
\documentclass[times]{qjrms4}
\usepackage[colorlinks,bookmarksopen,bookmarksnumbered]{}
\usepackage[draft]{hyperref}
\usepackage{amsmath}
\usepackage{amssymb}
\usepackage{upgreek}
\usepackage{bbm}
\usepackage{dsfont}
\usepackage{sansmath}
\usepackage{gensymb}
\usepackage{mathrsfs}
\usepackage[T1]{fontenc}

\input{chocros}

\begin{document}

\runningheads{I.~Polichtchouk and J.~Y-K.~Cho}{Superrotation by weak
  surface temperature gradient}

\title{Equatorial superrotation in Held \& Suarez-like flows\\ with
  weak equator-to-pole surface temperature gradient}

\author{I.~Polichtchouk\affil{a,b}\corrauth and J.~Y-K.~Cho\affil{b,c}$^\dagger$}

\address{ \affilnum{a} Department of Meteorology, University of
  Reading, Reading RG6 6BX, UK\\ \affilnum{b} School of Physics and
  Astronomy, Queen Mary University of London, London E1 4NS,
  UK\\ \affilnum{c} Institute for Theory and Computation, Harvard
  University, Cambridge, MA 02138, USA }

\corraddr{Department of Meteorology, University of Reading, Reading
  RG6 6BX, UK. E-mail: I.Polichtchouk@reading.ac.uk, \\
$^\dagger${\small On leave from Queen Mary University of London}}

\begin{abstract}

 Equatorial superrotation under zonally-symmetric thermal forcing is
 investigated in a setup close to that of the classic
 \cite{HeldSuarez94} setup. In contrast to the behaviour in the
 classic setup, a transition to equatorial superrotation occurs when
 the equator-to-pole surface equilibrium entropy gradient is
 weakened. Two factors contribute to this transition: 1)~the reduction
 of breaking Rossby waves from the mid-latitude that decelerate the
 equatorial flow and 2)~the presence of barotropic instability in the
 equatorial region, providing stirring to accelerate the equatorial
 flow. In the latter, Kelvin waves excited by instability near the
 equator generate and maintain the superrotation. However, the
 superrotation is unphysically enhanced if simulations are
 under-resolved and/or over-dissipated.
  
\end{abstract}


\maketitle

\section{Introduction}\label{sec1}

Superrotation is an important phenomenon in atmospheric flows.  It is
defined as a zonal-mean zonal flow with axial angular momentum, $M =
M(\phi)$, greater than the solid body angular momentum at the equator,
$M_0 = M(0)$; here $\phi$ is the latitude.  For a ``shallow''
atmosphere, $M = a\cos\phi(U + \Omega a\cos\phi)$ and therefore $M_0 =
\Omega a^2$, where $a$ is the planetary radius, $U$ is the zonal-mean
zonal flow and $\Omega$ is the planetary rotation rate.  Because
superrotation away from the equator is inertially
unstable\footnote{When superrotation is not present at the equator.},
it is mainly manifested as prograde (westerly) flow at the equator.
If friction maintains a solid body rotation at the surface (i.e., $M =
M_0\cos^2\phi$ at the surface), a zonally-symmetric mean meridional
circulation conserves $M$ such that $M \leq M_0$ at all latitudes in
the absence of sources and sinks. In general, superrotation can be
generated if there is a process which provides a steady eddy angular
momentum flux directed up the meridional gradient of the angular
momentum, $a^{-1} \del M_\phi/\del \phi$ \citep[][and see also Read
  1986]{Hide69}.

Superrotation is observed in the Earth's atmosphere (e.g. the
quasi-biennial oscillation), but it is only transient.  The current
explanation for the transience is that Rossby waves generated by
baroclinic instability at mid-latitudes propagate upwards, then
refract equatorward and break in the equatorial region
\citep[e.g.][]{Held85}.  Because the waves deposit prograde momentum
in the source region and retrograde momentum in the breaking or
saturation region, equatorial superrotation does not occur.  On the
other hand, superrotation does occur more robustly on other planets
and in idealized general circulation model (GCM) experiments.  For
example, in the latter, superrotation occurs under a variety of
forcing conditions: zonally-asymmetric tropical heating
\citep[e.g.][]{Suarez92,Saravanan93,Hoskins99,Arnold12},
zonally-symmetric heating with a convective equatorial wave source
\citep[e.g.][]{Schneider09,Laraia15}, and statically-stable
zonally-symmetric heating with $\Omega$ or $a$ smaller than those of
the Earth \citep[e.g.][]{Williams88,Williams03,Mitchell10}. In the
statically-stable zonally-symmetric case, the increase in the global
scale Rossby number --- effected by reducing $\Omega$ or $a$ ---
results in a weakened mid-latitude baroclinic instability and a
strengthened barotropic instability.  \cite{Williams03} and
\cite{Mitchell10} have attributed superrotation to barotropic
instability.

\cite{Williams03} has also shown that superrotation
can occur on a planet with Earth's $\Omega$ and $a$ subject to
statically-stable zonally-symmetric heating, when the center of
maximum baroclinicity is shifted to a lower latitude than that in the
classic setup of \citet{HeldSuarez94} [hereafter HS].  The shift is
effected by narrowing the meridional width of the radiative heating
profile compared to that in HS. In this case the barotropic
instability at the equatorward flank of the subtropical jet provides
the mechanism for producing superrotation. \cite{Williams06} has
extended this work by showing that high tropospheric static stability
enhances superrotation by inhibiting baroclinic instability.

Recently the barotropic instability view has been questioned by
\cite{potter2014}. They propose that equatorial Kelvin waves drive and
maintain superrotation in small $\Omega$ or $a$ flows and in
\cite{Williams03}, although the origin and precise driving mechanism
of the Kelvin waves were not identified in their study.  Later,
\cite{Wang14} attribute the Kelvin waves to a linear
resonant coupling of equatorial Kelvin and higher-latitude Rossby
waves originating from an ageostrophic instability
\citep[e.g.][]{Sakai89}: the instability `funnels' westerly momentum
to the equator, producing superrotation. However, as we shall show, in
our study the superrotation is likely generated and maintained by
different mechanisms.

Our study can be regarded as an extension of \cite{Williams03,
  Williams06}, \cite{Mitchell10}, \cite{potter2014} and \cite{Wang14}.
Here transition to superrotation is also investigated under
statically-stable zonally-symmetric thermal forcing; however, it is
shown that superrotation also occurs with Earth's $\Omega$ and $a$, if
the equator-to-pole equilibrium temperature gradient near the surface
(from 1000\,hPa to 700\,hPa) is weakened, compared to that in
HS. While such a setup is not necessarily more (or less) realistic, it
does help to elucidate the processes involved in producing
superrotation. We show that when the temperature gradient is weak,
baroclinic instability is weakened and barotropic instability in the
equatorial region excites Kelvin waves, which help to drive
superrotation. The key ingredient in the excitation is the stirring
provided by the instability that occurs within the equatorial
deformation length scale of a Kelvin wave. Further, we show that the
strength of the resulting superrotation is strongly affected by
numerical resolution and dissipation: superrotation is artificially
strengthened by explicit and implicit model diffusion.

Throughout this paper, the reduced temperature gradient region is
denoted as the `RTG' region.  Such a reduced gradient region is not
necessarily meant to address the present-day Earth, but it may be
relevant for past and future climates of the Earth or other
`Earth-like' planets in the Solar and extrasolar systems. For example,
such a thermal forcing can be applicable to the runaway greenhouse
state of paleo-Earth or the present-day atmosphere of Venus, both of
which are characterized by a weak equator-to-pole surface temperature
gradient \citep[see e.g.][]{Tziperman09,Yamamoto03, Lee05}.

The overall plan of the paper is as follows. In section~\ref{sec2} the
primitive equation model and the setup used in this study is
discussed.  In section~\ref{sec3} results from the model simulations
with a RTG region are presented and compared with the classic HS-like
simulations.  This section also discusses in detail the
non-convergence of simulations when they are under-resolved or
over-dissipated. Finally, conclusions are given in section~\ref{sec4}.

\section{Method}\label{sec2}

\subsection{Numerical model}
In this study a global pseudospectral model, BOB \citep{Scott03}, is
used. BOB solves the `dry' hydrostatic primitive equations in pressure
($p$) coordinates with rigid boundary conditions at the top and bottom
of the model domain.  The equations are solved in vorticity-divergence
form.  A parallel pseudospectral algorithm is used in the horizontal
direction and a second-order finite differencing is used in the
vertical direction.  A hyperviscosity operator $-\nu\nabla^8$, with
constant $\nu$, is applied to the prognostic variables (relative
vorticity $\zeta$, divergence $\delta$ and potential
temperature~$\theta$). The timestepping is performed using a
semi-implicit, second-order leapfrog scheme and the Robert-Asselin
filter \citep{Robert66,Asselin72} with a small filter coefficient
($=\! 0.02$) is applied at each timestep to stabilize the time
integration.

The physical setup is essentially that of HS, as well as
\citet{Williams03} (his case $A$).  A simple Newtonian relaxation
scheme for the net heating,
\begin{equation} 
  {\cal{Q}} = -\frac{1}{\tau}\Big(\theta-\theta_{\text{e}}\Big),
\end{equation}
is applied in the potential temperature tendency equation; here $\tau$
is the characteristic relaxation time, and $\theta_{\text{e}} =
\theta_{\text{e}}(\phi,p)$ is a specified equilibrium potential
temperature distribution.  The surface friction is parametrized as a
linear, Rayleigh drag in the momentum equation:
\begin{equation} 
  {\cal{D}}\ =\ -\frac{{\bf v}}
  {\tau_{\text{\tiny R}}}\ \max\!\left\{0,\,\frac{p-p_{\text{b}}}
    {p_{\text{s}} - p_{\text{b}}}\right\},
\end{equation}
where ${\bf v}$ is the flow on an isobaric surface, $\tau_{\text{\tiny
    R}}$ is the characteristic damping time and $p_{\text{b}}$ and
$p_{\text{s}}$ are the pressures at the top of the boundary layer
(700\,hPa) and at the bottom surface (1000\,hPa), respectively.  The
only important physical difference relevant for superrotation in this
setup, compared to that of HS, is $\theta_{\text{e}}$.  This is discussed
more in detail below.

\begin{table}\small
  \caption{Parameter values used in the simulations: $g$ is the
    surface gravity; $a$ is the radius; $\Omega$ is the rotation rate;
    ${\cal R}$ is the gas constant; $c_p$ is the specific heat at
    constant pressure; $p_{\text{s}}$ is the surface pressure;
    $p_{\text{b}}$ is the top of the boundary layer pressure; $\tau$
    is the radiative relaxation time; $\tau_{\text{\tiny R}}$ is the
    Rayleigh drag time; $ T_{\text{s}}$ and $T_0$ are the
    stratospheric and upper tropospheric reference temperatures,
    respectively; $\Delta T$ is the equator-to-pole temperature
    difference; and, $\Delta\theta$ is the vertical potential
    temperature difference.}  \centering
\begin{tabular}{llll}
\toprule
 Parameter \hspace*{3mm} & Value \hspace*{1cm} & Units \\
\midrule

$g$ & 9.8 & m~s$^{-2}$ \\ 
$a$ & $6.4\!\times\! 10^{6}$ & m\\ 
$\Omega$ & $7.3\!\times\! 10^{-5}$ & s$^{-1}$ \\
${\cal R}$ & $287$ & J~kg$^{-1}$~K$^{-1}$ \\ 
$c_{p}$ & $1004$ & J~kg$^{-1}$~K$^{-1}$\\ 
$p_{\text{s}}$ & $1000$ & hPa \\
$p_{\text{b}}$ & $700$ & hPa \\
$\tau$ & $20$ & days \\
$\tau_{\text{R}}$ & $1$ & day \\
$ T_{\text{s}}$ & $200$ & K \\
$ T_0$ & $255$ & K \\
$\Delta T$ & $60$ & K \\
$\Delta \theta$ & $10$ & K \\
\\ \bottomrule
\end{tabular}\label{table1}
\end{table}

\begin{figure*}
\centering
  \includegraphics[width=\textwidth]{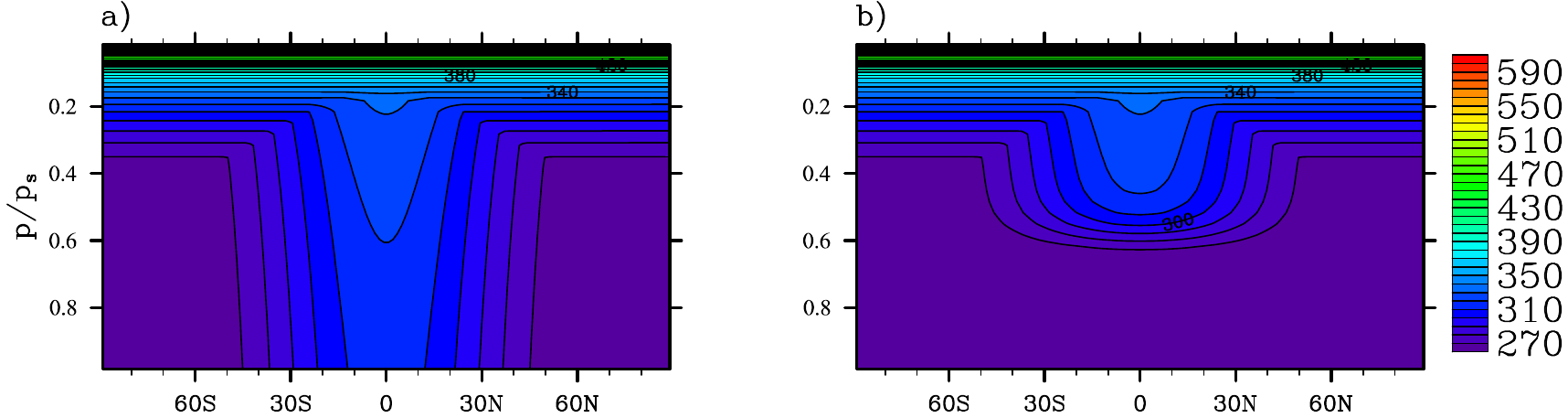}
  \caption{Prescribed zonally-symmetric radiative
      equilibrium potential temperature distribution
      $\theta_{\text{e}}$\,[K], normalized pressure vs. latitude, for
      the control (HS-like) case ($a$) and the reduced temperature
      gradient (RTG) case ($b$).  The contour interval is
      10\,K.}\label{fig1}
\end{figure*}

\subsection{Setup}\label{setup}
The physical parameters used in all the simulations discussed in this
study are listed in Table~\ref{table1}.  The simulations are
initialized with an isothermal state of rest, with a small amount of
Gaussian white noise ($10^{-7}$\,s$^{-1}$ standard deviation)
introduced to the $\zeta$ field to break symmetry.  The specified
equilibrium potential temperature is
\begin{align}\label{eqT}
  \theta_{\text{e}}(p,\phi) = \text{max} \Biggl\{ T_{\text{s}}\left(\frac{p_{\text{s}}}{p}\right)^\kappa,
  \Bigl[ T_0 + \Delta T\,\cos^{b}\!\phi~F(p) \nonumber \\ 
  -
 \Delta \theta\,\log\left(\frac{p}{p_s}\right) \cos^2\!\phi\Bigr]
   \Biggr\}\, . 
\end{align}
Here $T_{\text{s}}$ and $T_0$ are the stratospheric and upper
tropospheric reference temperatures, respectively; $\Delta T$ is the
equator-to-pole temperature difference; $\Delta \theta$ is the
vertical potential temperature difference; and, $\kappa \equiv R/c_p =
2/7$.

In equation~(\ref{eqT}), $b$ and $F(p)$ control the meridional width
of the equilibrium temperature and the equator-to-pole temperature
gradient near the surface, respectively; the masking function $F$ is
defined,
\begin{equation}\label{Fp}
  F(p)\ =\ \exp\left\{-\frac{(p/p_{\text{s}})^{12}}{ \gamma^2}\right\},
\end{equation}
where $\gamma$ is a thickness parameter and is used to specify the
vertical `$e$-folding' extent of the RTG region.  With $b = 4$ and
$F(p) = 1$ (i.e. $\gamma \rightarrow \infty$), $\theta_{\text{e}}$ reduces
to the HS case, which roughly captures the statistical mean state of
the Earth's troposphere.  The resulting $\theta_{\text{e}}$ distribution is
shown in Figure~\ref{fig1}$a$.  Here the magnitude of the
equator-to-pole equilibrium temperature gradient in the lower
troposphere is $\approx\!\Delta T$.  In section~\ref{sec3} simulations
with $\gamma = 0.045$ are mainly discussed, for which $\theta_{\text{e}}$
is shown in Figure~\ref{fig1}$b$.  In the present study, the focus is
on the influence of $\gamma$ on superrotation; however, we note that
the generation and strength of superrotation are highly sensitive to
all of $\{\gamma, b, \Delta T, T_{\rm s}\}$, in general.  This broad
sensitivity to the physical parameters is discussed in
section~\ref{sec3.4}.

\begin{figure*}
 \centering
\includegraphics[width=\textwidth]{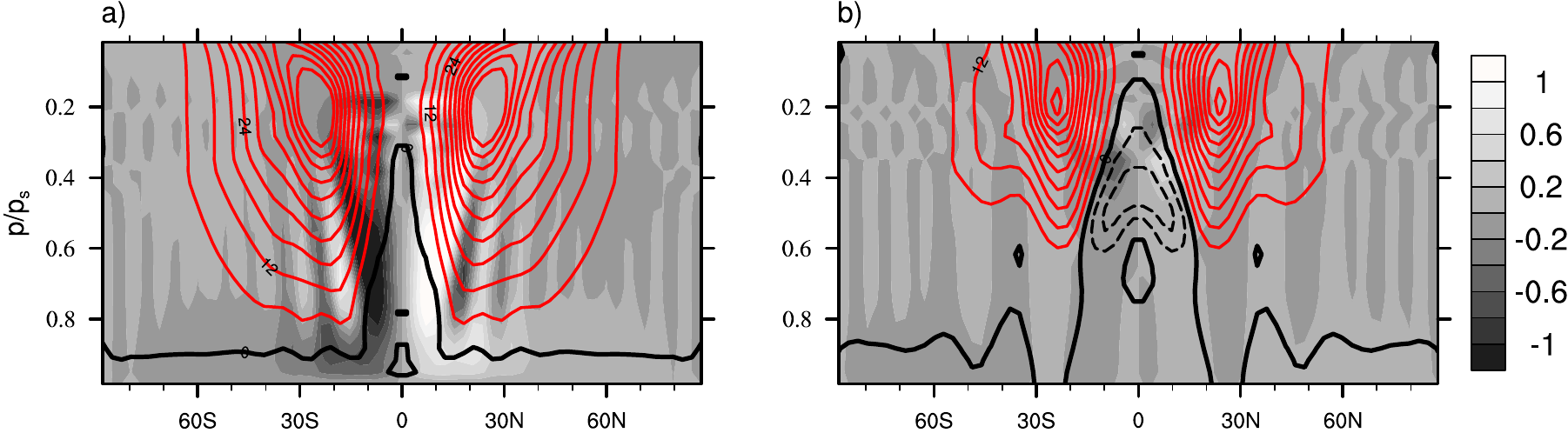}\\ 
\caption{The axisymmetric zonal-mean zonal wind
  $\overline{u}$\,[m\,s$^{-1}$] (contours) and mass streamfunction
  $\overline{\psi}\times 3 \times 10^{-11}$\,[kg\,s$^{-1}$] (in
  greyscale) for the control (HS-like) ($a$) and RTG ($b$) simulations
  at T170L30 resolution, with corresponding $\theta_e$ from
  Figure~\ref{fig1}. Maximum and minimum values for $\overline{u}$ are
  $\pm$\,60\,m\,s$^{-1}$, with contour interval
  6\,m\,s$^{-1}$. Negative $\overline{u}$ values are dashed (and in
  black) and positive are solid (and in red). The zero contour is drawn
  with double thickness. The light shading corresponds to the
  clockwise circulation and the dark shading corresponds to the
  anticlockwise circulation, respectively. No superrotation is present
  in the axisymmetric, eddy-free case.  }\label{fig2add}
\end{figure*}
\begin{figure*}
\hspace{-0.5cm} 
\centering 
$\begin{array}{cc}
  \includegraphics[width=\textwidth]{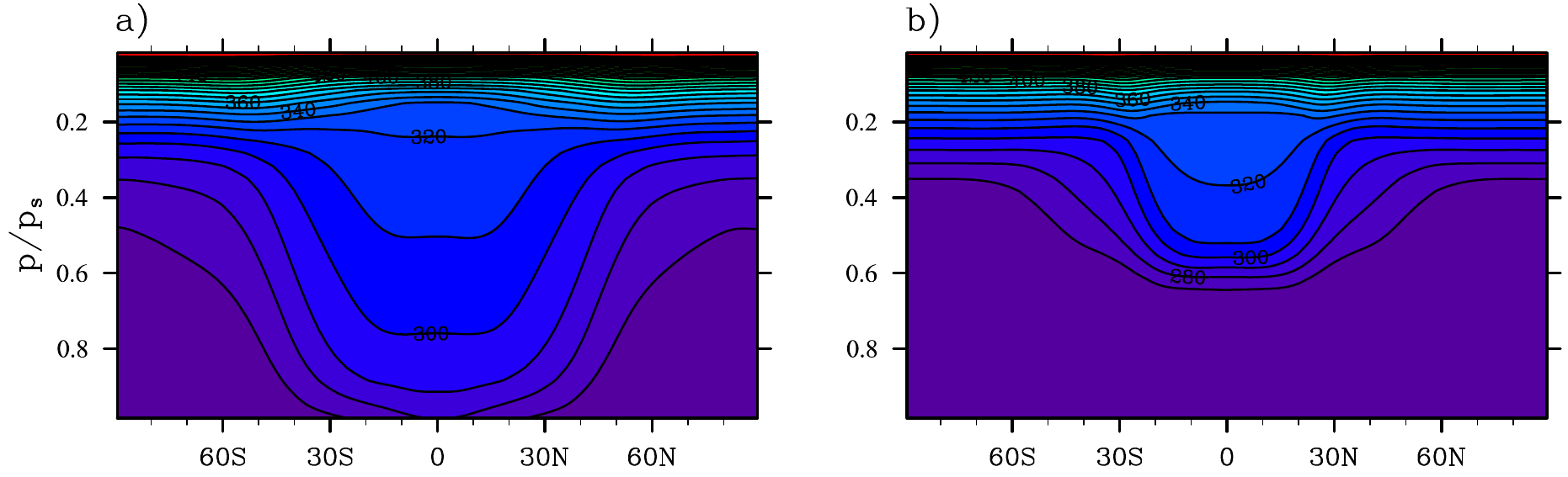}\\ 
  \includegraphics[width=\textwidth]{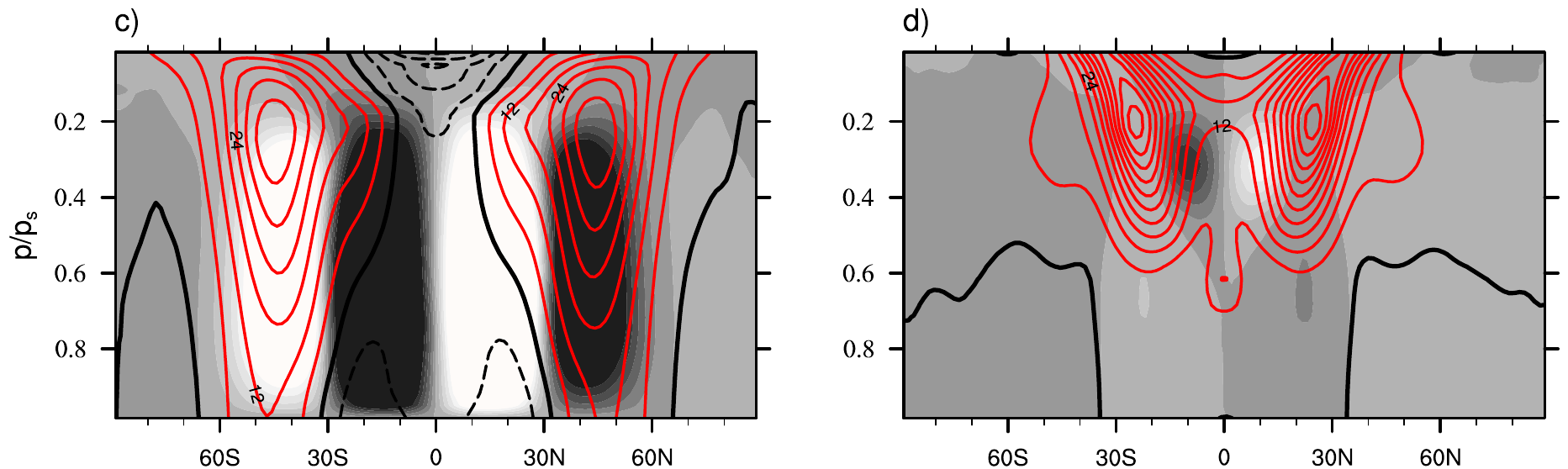}  
\end{array}$\\
\caption{The time- and zonal-mean potential temperature
  $\overline{\theta}^*$[K], zonal wind
  $\overline{u}^*$\,[m\,s$^{-1}$], and mass stream function
  $\overline{\psi}^* \times 3 \times 10^{-11}$\,[kg\,s$^{-1}$] for the
  fully turbulent three-dimensional control (HS-like) ($a$\,and\,$c$)
  and RTG ($b$\,and\,$d$) simulations at T170L30 resolution, with
  corresponding $\theta_e$ from Figure~\ref{fig1}.  The time averages
  are taken over $t=[700,1200]$~days ($a$\,and\,$c$), and over
  $t=[4500,5000]$~days ($b$\,and\,$d$). Contour values for
  $\overline{\theta}^*$ are the same as in Figure~\ref{fig1} and
  values for $\overline{u}^*$ and $\overline{\psi}^*$ are the same as
  in Figure~\ref{fig2add}. Clear superrotation is generated at the
  equator (at all altitudes) in the fully turbulent RTG simulation.
}\label{fig2}
\end{figure*}

Note also that, apart from $F$, there are additional differences in
the setup used in HS, \cite{Williams03} and the present study.  For
example, in \cite{Williams03} $\Delta \theta$ in equation~(\ref{eqT})
is not modulated by $\cos^2\!\phi$.  In addition, $\tau = 20$~days
everywhere in this study -- different than in HS, but the same as in
\cite{Williams03}.  None of these differences qualitatively affect the
results presented here, however.

The model domain, which extends vertically from $10^3$\,hPa to
$17$\,hPa, is resolved by 30 levels equally spaced in $p$. Simulations
are performed at \{T42, T85, T170, T341\} horizontal resolutions.  The
`control simulations' (i.e. essentially HS simulations), with
$\theta_{\text{e}}$ shown in Figure~\ref{fig1}$a$, are integrated for
$t = 1200$\,days; and, the `RTG simulations', with $\theta_{\text{e}}$
shown in Figure~\ref{fig1}$b$, are integrated for up to $t =
7000$\,days.  In the latter simulations, equilibration takes longer at
lower horizontal resolutions and/or with stronger dissipation.  The
timestep size $\Delta t$ (in seconds) is \{600, 300, 150, 75\} for the
resolutions given above, respectively.  The hyperdiffusion coefficient
$\nu$ is chosen so that the $e$-folding time $\tau_d$ (in days) for
the smallest scale in the system is \{0.1, 0.01, 0.001, 0.0001\},
respectively.  Assuming
\begin{equation}
  \tau_d\ \approx\ \frac{1}
  {\nu}\left[\frac{a^2}{N_{\mbox{\tiny T}}\,(N_{\mbox{\tiny T}} + 1)}\right]^4,
\end{equation}
where $N_{\mbox{\tiny T}}$ is the spectral truncation wavenumber, the
above $\tau_d$'s correspond to $\nu$ (in $10^{37}$\,m$^8$\,s$^{-1}$)
of \{3.0, 0.1, 0.004, 0.0002\}, respectively.  Sensitivity of the
results to horizontal resolution and hyperdissipation specifications
is discussed in section~\ref{sec3.3}.

\section{Results}\label{sec3}

\subsection{Mean circulation}\label{sec3.1}

Before describing the turbulent three-dimensional simulations, it is
fruitful to examine the axisymmetric eddy-free circulation that
results from prescribing the $\theta_e$ shown in Figure~\ref{fig1} and
spinning up the atmosphere from an initially resting state.
Figure~\ref{fig2add} shows the resulting axisymmetric zonal-mean zonal
wind and mass streamfunction ($\overline{u}$ and $\overline{\psi}$,
respectively) from the control ($a$) and RTG ($b$) simulations at
T170L30 resolution (i.e. T170 horizontal resolution with 30 vertical
levels).  The axisymmetric circulations are obtained by running the
BOB model with no initial perturbation\footnote{The axisymmetric
  simulations are integrated for $t\approx$500\,days, up until gravity
  waves degrade the flow.}. Both simulations produce a weak Hadley
circulation and the zonal flow consists of weak subrotation (easterly
flow) at the equator and westerly jets centred at $\phi \approx \pm 25
\degree$.  However, the final axisymmetric circulation is dependent on
the initial condition, in general.

Figure~\ref{fig2} shows the time- and zonal-mean potential
temperature, zonal wind and mass streamfunction meridional
cross-sections ($\overline{\theta}^*$, $\overline{u}^*$ and
$\overline{\psi}^*$, respectively) from the fully turbulent,
three-dimensional control ($a$\!~and~\!$c$) and RTG ($b$\!~and~\!$d$)
simulations. Time-averaging is performed over $25\,\tau$'s after the
simulations have reached a statistically equilibrated state, over the
intervals $t~\!=\!~[700, 1200]$\,days for the control simulation and
$t\!~\!=\!~[4500, 5000]$\,days for the RTG simulation.
\begin{figure*}
\centering $ \begin{array}{ccc}
  \includegraphics[width=0.5\textwidth]{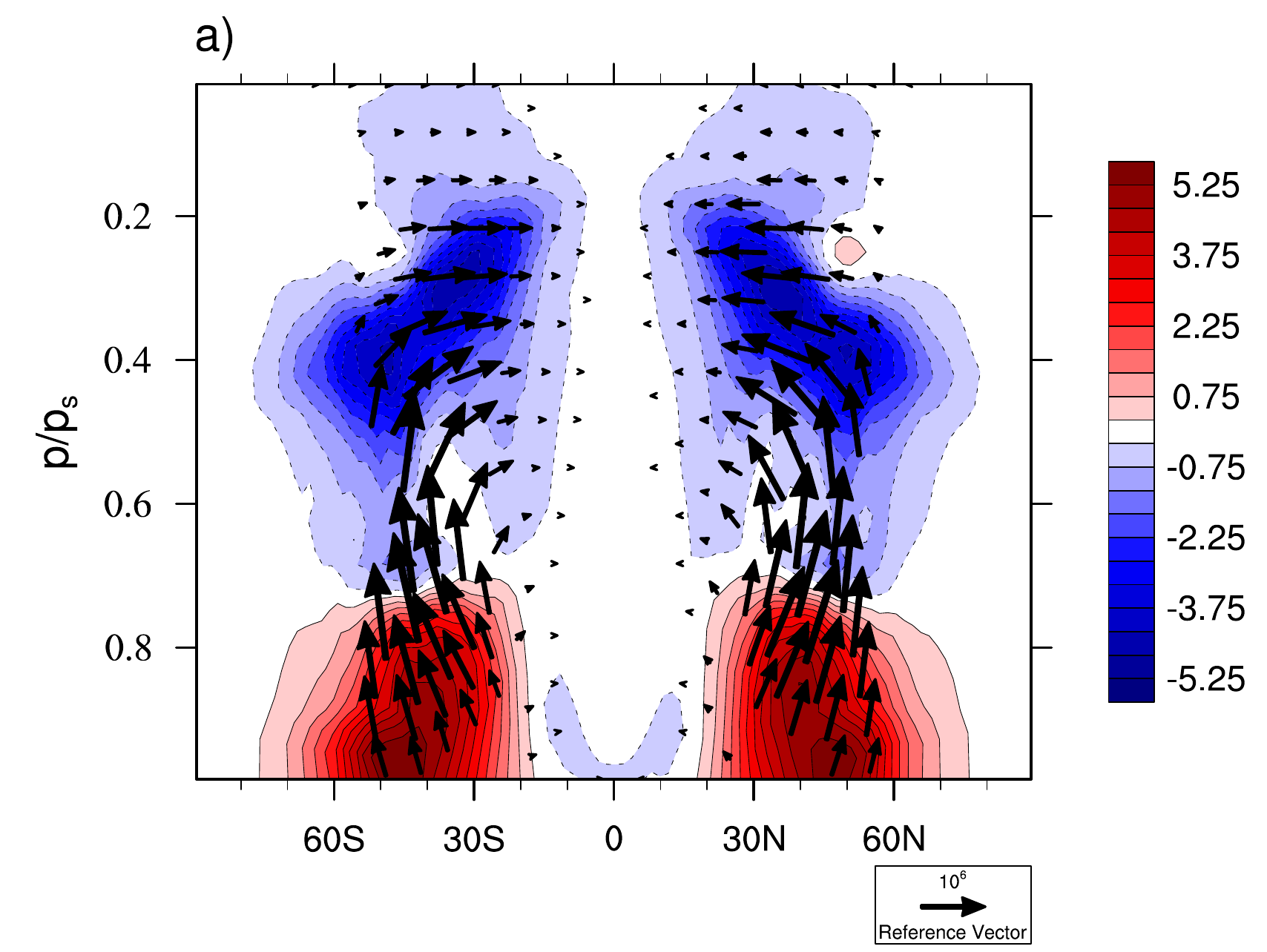}
  \includegraphics[width=0.5\textwidth]{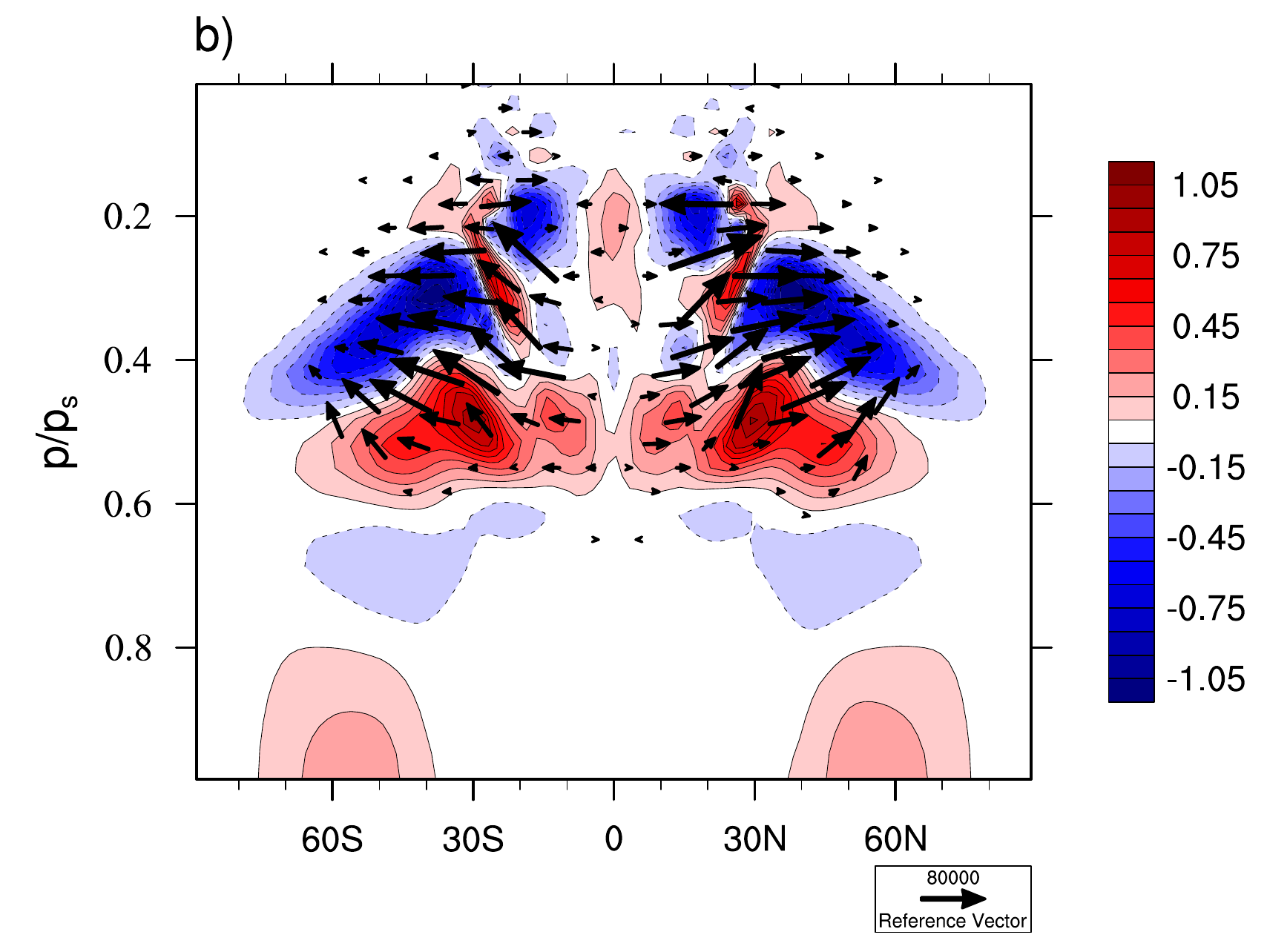}
  \end{array}$\vspace{-0.3cm}
  \caption{Eliassen-Palm flux vector, ${\bf F} = (F_{\phi},
    F_p)$\ [m$^2$~s$^{-2}$], and its divergence, $\nabla\! \cdot\!
    {\bf F}$\ $\times 10^9$ [m~s$^{-2}$], fields for the control~($a$)
    and RTG~($b$) simulations at T170L30 resolution averaged in time
    during the equilibrated stage.  Negative $\nabla\!\cdot\!{\bf F}$
    values are dashed (and in blue) and positive values are solid (and
    in red). $F_{\phi}$ and $F_p$ are scaled by 1~radian of latitude
    and 1~Pa of pressure, respectively.  }\label{fig3}
\end{figure*}

In the control simulation, initially zonally-symmetric westerly jets
(cf. Figure~\ref{fig2add}$a$) become baroclinically unstable and
zonally-asymmetric flow develops.  Following the onset of baroclinic
instability, the jets move poleward and equilibrate at $\phi \approx
\pm 45\degree$ (Figure~\ref{fig2}$b$).  The exact mechanism leading to
the poleward migration is currently not well understood, but all of
the proposed mechanisms involve some form of eddy-mean flow
interaction \citep[see e.g.][]{Chen07,Kidston10}.  Crucially, the flow
remains weakly easterly at the equator, at all heights in the control
simulation. Baroclinic eddies transport heat from low to high
latitudes; as a result, the equator-to-pole temperature gradients are
weaker in the equilibrated $\overline{\theta}^*$ distribution than in
the prescribed $\theta_{\rm e}$ distribution (cf. Figure~\ref{fig2}$a$
with Figure~\ref{fig1}$a$). Note that the Hadley circulation in the
fully non-linear simulation is mainly eddy-driven
(cf. Figures~\ref{fig2add}$a$ and \ref{fig2}$c$).

In contrast, the equilibrated flow in the RTG simulation is
superrotating at the equator when the eddies are present -- and there
at all altitudes, except at the very top (cf. Figures~\ref{fig2add}$b$
with Figure~\ref{fig2}$d$). There is also a weak Hadley circulation in
the RTG simulation in this case. Accordingly, the
$\overline{\theta}^*$ distribution changes only slightly from the
prescribed $\theta_{\rm e}$ distribution (cf. Figure~\ref{fig2}$b$
with Figure~\ref{fig1}$b$).  This is consistent with the marked
reduction in baroclinic instability observed in this simulation. The
equilibrated subtropical jets remain close to their initial latitude
(i.e. at $\phi \approx \pm 25\degree$) and take a much longer time to
equilibrate, compared to the control simulation. The total
(column-mean) equatorial zonal wind $\langle u_{\rm eq} \rangle$
undergoes rapid acceleration over $t = [200, 1000]$\,days, reaching
statistical equilibration thereafter. Throughout this paper, this
stage is referred to as the `equilibrated stage' and the stage prior
to this as the `acceleration stage'.

\subsection{Wave-mean flow analysis}\label{sec3.2}

Acceleration of the zonal-mean zonal flow and direction of wave
propagation can be assessed via the Eliassen-Palm (EP) flux vector,
${\bf F }= (F_{\phi}, F_{p})$ \citep[e.g.][]{Andrews78}:
\begin{eqnarray}
  F_{\phi} & = & a
  \cos\phi\left[-\overline{u'v'} + 
    \left(\frac{\overline{v'\theta'}}{\del \overline{\theta}/ \del p}\right)\del \overline{u}/\del p \right] \label{EP1}\\ 
  F_p & = & a \cos\phi
  \left[(\overline{\zeta} + f)
    \left(\frac{\overline{v'\theta'}}{{
          \del \overline{\theta} /\del p}}\right)
    -\overline{u'\omega'}\right] \label{EP2}, \
\end{eqnarray}
where the primes denote deviations from the zonal mean, $v$ is the
meridional velocity, $f = 2\Omega \sin\phi$ is the Coriolis parameter,
$\theta \equiv T (p/p_s)^{\kappa}$ is the potential temperature and
$\omega = Dp/Dt$ is the `vertical' velocity.  The second term in the
brackets in equation~(\ref{EP1}) and the first (i.e. the
$\overline{\zeta}$ term) and third terms in equation~(\ref{EP2}),
which are the ageostrophic terms, are generally small compared to the
remaining, geostrophic terms.

Figure~\ref{fig3} shows the EP flux vector field and its divergence,
$\nabla\!\cdot\!{\bf F}$ (contours), for the control~($a$) and
RTG~($b$) simulations.  In the control simulation, the EP flux and its
divergence follow the classic scenario: Rossby waves, generated near
the surface by mid-latitude baroclinic instability, transport wave
activity upward in the troposphere and equatorward at mid-height,
transporting momentum poleward and heat upward \emph{and} poleward.
In the northern hemisphere, according to equation~(\ref{EP1}),
$F_{\phi} < 0$ implies $\overline{u'v'} > 0$ (assuming the
ageostrophic term to be small), leading to a poleward eddy momentum
flux; correspondingly, according to equation~(\ref{EP2}), $F_p > 0$
implies $f\overline{v'\theta'} > 0$, leading to a poleward eddy heat
flux. The vector field is divergent (positive) near the surface at
mid- and high-latitudes and convergent (negative) in the upper
troposphere and lower stratosphere, as well as in the equatorial
region near the surface. According to the \emph{transformed Eulerian
  mean} (TEM) equations \cite[e.g.][]{Andrews87}, zonal flow is
accelerated (decelerated) in regions of positive (negative)
$\nabla\!\cdot\!{\bf F}$; hence, the jet becomes increasingly
barotropic.

In the RTG simulation, the transport of wave activity is upward and
mainly poleward, unlike in the control simulation (see
Figure~\ref{fig3}$b$). The poleward heat flux is considerably weaker
(n.b. the smaller reference vector magnitude in Figure~\ref{fig3}$b$)
and the flux activity is clearly dominant at higher altitudes (above
$p \approx 600$\,hPa level).  The eddy momentum flux is mostly
equatorward in both the northern and southern hemispheres. The
$\nabla\!\cdot\!{\bf F}$ field is also weaker than in the control
simulation (n.b. the smaller $\nabla\!\cdot\!{\bf F}$
  values in Figure~\ref{fig3}$b$) and displays a more complicated
structure than in the control simulation. Significantly, a westerly
zonal flow acceleration is present in the tropics, in the
$p\approx[250, 600]$\,hPa region. Note also that at $p=200$\,hPa, the
zonal flow is accelerated at the equator and decelerated in the
subtropics, in the $10\degree<|\phi|<30\degree$ region, where the wave
propagation is equatorward. In addition, the poleward flanks of the
subtropical jets are decelerated at higher altitudes ($p \approx [200,
  450]$\,hPa) and accelerated at lower altitudes ($p \approx [450,
  600]$\,hPa).

\begin{figure*}
\centering 
$\begin{array}{cc}
  \includegraphics[width=0.5\textwidth,height=14cm]{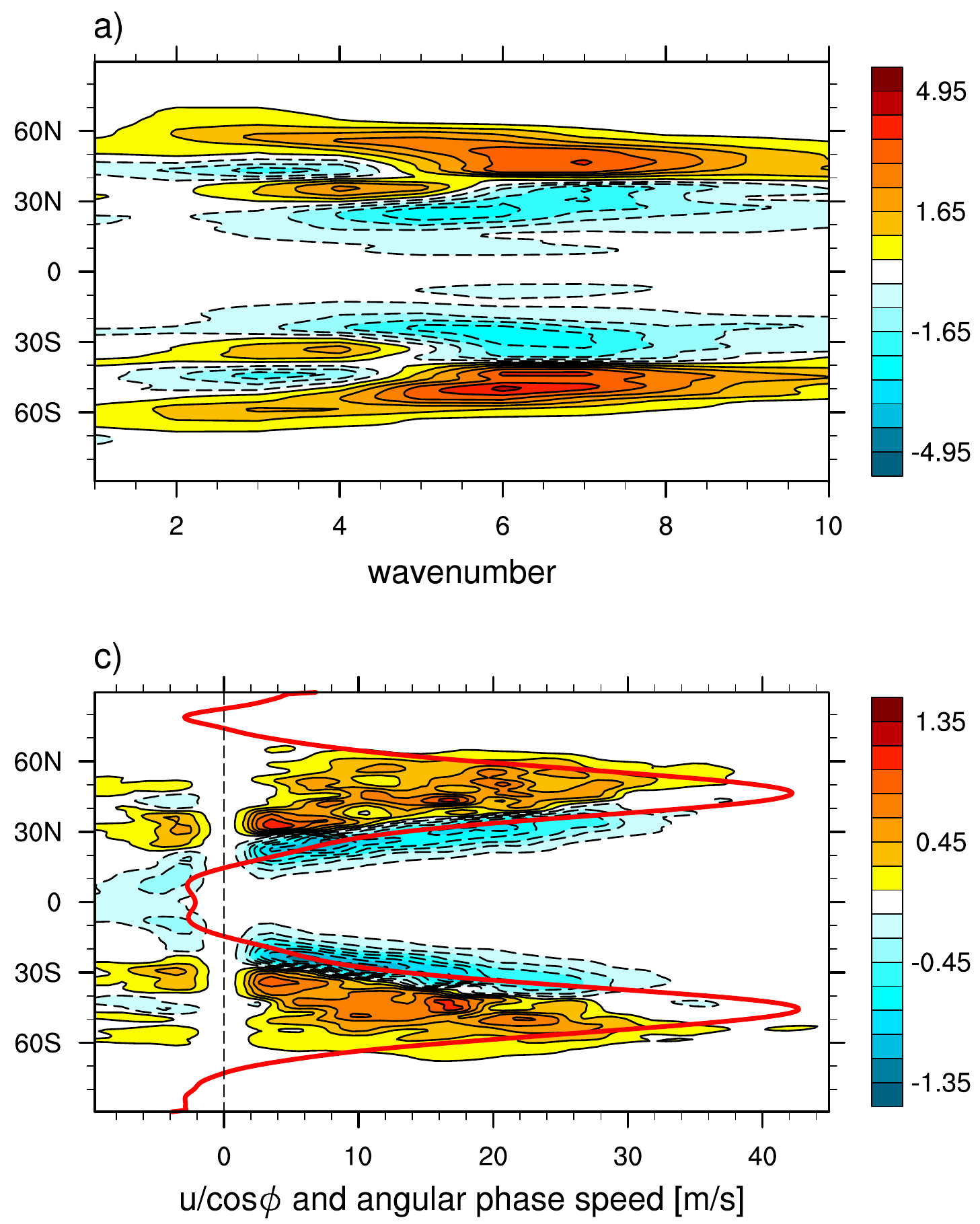}\hspace{0.7cm}
  \includegraphics[width=0.5\textwidth,height=14cm]{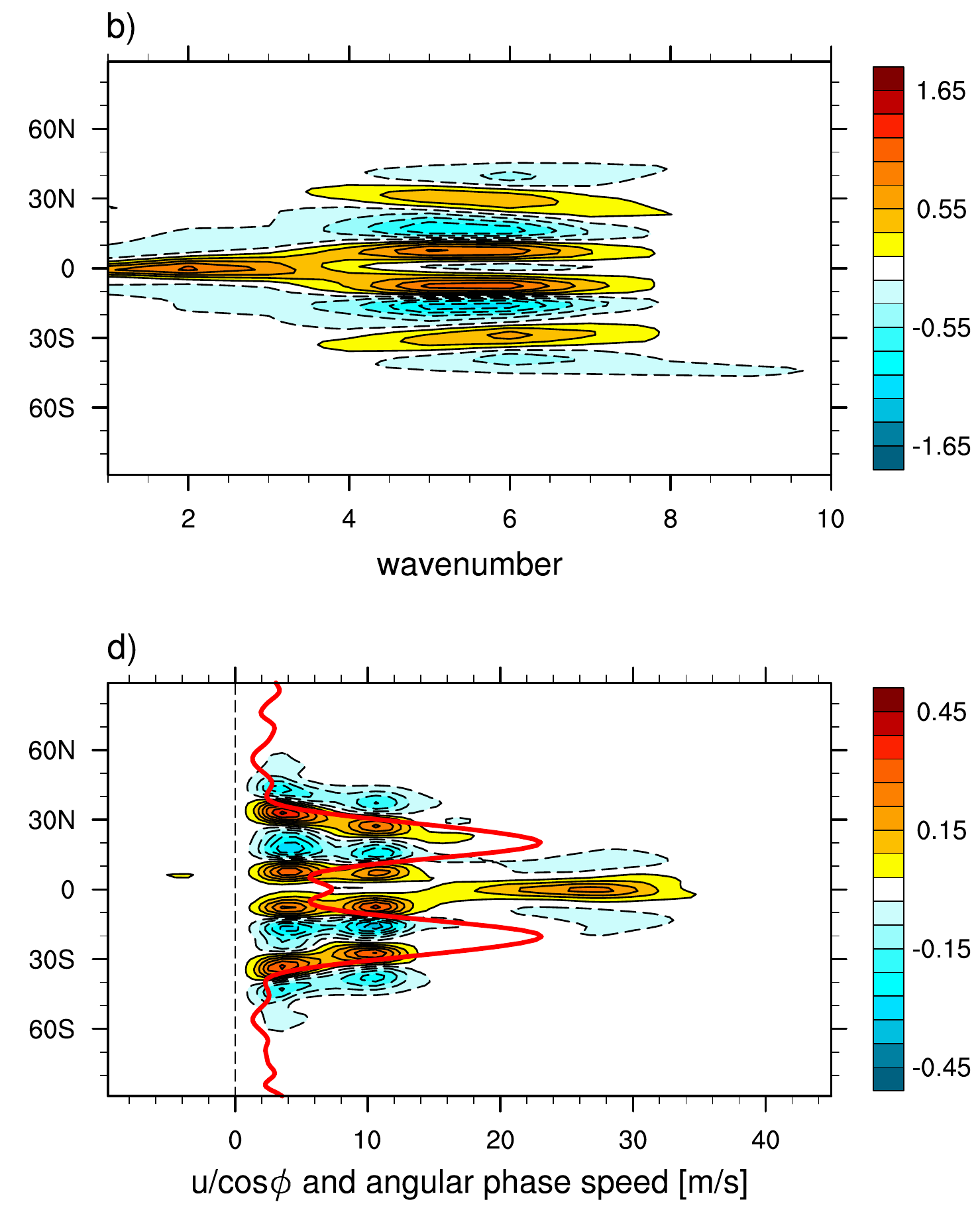}  
\end{array}$\\
       \caption{Eddy momentum flux convergence spectra $\times
         10^{-6}$\,[m\,s$^{-2}$] at $p = 500$\,hPa for the control and
         RTG simulations at T170L30 resolution: ($a$) and ($b$) show
         the $m$--$\phi$ spectrum and ($c$) and ($d$) show the
         $c_A$--$\phi$ spectrum (contours) as well as the angular
         velocity $\overline{u}/\cos\phi$ (red curve).  The spectra
         are computed from ten 50-day windows over $t =
         [700,1200]$\,days for the control simulation ($a$ and $c$)
         and over $t = [4500,5000]$\,days ($b$ and $d$) for the RTG
         simulation.  Negative contours are dashed and positive
         contours are solid.  Notice the dominance of zonal
         wavenumbers $m \approx 6$ in both simulations.  The phase
         speeds in the control simulation have values less than or
         comparable to $\overline{u}/\cos\phi$\ins{,} whereas
         $c_A>\overline{u}/\cos\phi$ in the RTG simulation at the
         equator.  Critical layers are located at latitudes where $c_A
         \approx \overline{u}/\cos\phi$.  High phase speed, zonal
         wavenumber $m\!\approx\!  2$ modes help to accelerate the
         zonal mean zonal flow at the equator in the RTG simulation.
       }\label{fig5}
\end{figure*}

It is useful to examine the spectral characteristics of the eddy
momentum flux convergence.  For this, the eddy momentum flux at each
$\phi$ is first spectrally decomposed among the zonal wavenumbers
$\{m\}$ and the $m$--$\phi$ covariance spectrum is obtained before
computing the divergence \citep[e.g.][]{Hayashi71,Randel91}.
Similarly, the angular phase speed, $c_A\,(m,\phi) = \varpi a /m$ with
$\varpi = \varpi(m)$ the frequency, is computed to obtain the
$c_A$--$\phi$ spectrum. We use $c_A$ instead of phase speed $c$,
because $c_A$ is conserved following a meridionally propagating Rossby
wave packet in a zonally-symmetric background flow.  Note that a
positive (negative) region in the convergence spectrum is associated
with a momentum source (sink).  The convergence spectra from both the
control and RTG simulations are shown in Figure~\ref{fig5}.  Spectra
are taken from the $p = 500$\,hPa level, the level of maximum zonal
wind acceleration in the RTG simulation (see
Figure~\ref{fig3}$b$)\footnote{This level also lies just above the
  tropospheric static stability maximum.}.  Recall that there is no
superrotation in the control case.

First we discuss the $m$--$\phi$ spectra.  Figure~\ref{fig5}$a$ shows
the $m$--$\phi$ spectrum calculated for ten consecutive 50-day windows
over $t = [700, 1200]$\,days from the control simulation during the
equilibrated stage. The spectrum for only $m \leq 10$ is shown because
larger wavenumbers possess insignificant amplitudes.  The spectrum
shows that the modes of $m\!  \approx\!  6$, centered on the jets,
dominate at mid-latitudes. Figure~\ref{fig5}$b$ shows the
corresponding $m$--$\phi$ spectrum from the RTG simulation in the
equilibrated stage ($t = [4500, 5000]$\,days).  The spectrum in this
simulation is also dominated by $m\!\approx\! 6$ modes, but the waves are
excited on both flanks of the subtropical jets, at $\phi \approx \pm
5\degree$ and $\phi \approx \pm 35\degree$.  These waves stir the flow
and converge westerly momentum into their source region and easterly
momentum into their breaking/saturation region.  In addition, low
wavenumber (i.e. $m\!\approx\! 2$) modes converge westerly momentum at
the equator.  Later we show that the $m\!\approx\! 2$ modes have a
distinct Kelvin wave-like behaviour, as also identified by
\cite{potter2014} in their study.

There are other important, distinguishing features in the RTG
simulations.  For example, at lower model resolutions the convergence
spectra exhibit significant qualitative differences, compared to the
spectrum in Figure~\ref{fig5}$b$.  At resolutions lower than T170 (and
assuming the nominal values for the hyper-dissipation coefficients
(section~\ref{sec3.3})) the $m \!\approx\! 6$ modes, which are located
on the flanks of the subtropical jet, disappear completely after the
initial acceleration stage.  In the simulation presented in
Figure~\ref{fig5}$b$, these modes persist well into the equilibrated
stage.  As will be shown in section~\ref{sec3.3}, the disappearance is
due to inadequate horizontal resolution.

Now we discuss the $c_A$--$\phi$ spectra.
Figures~\ref{fig5}$c$\!~and~\!\ref{fig5}$d$ present the $c_A$--$\phi$
spectra at the $p = 500$\,hPa level for the control ($c$) and RTG
($d$) simulations, respectively.  Note that in both spectra, there is
insignificant amplitude at $c_A\leq-10~$m$\,$s$^{-1}$ and $c_A \ge
45$\,m\,s$^{-1}$.  The spectra are obtained from the data over the
same time windows as in the corresponding spectra in
Figures~\ref{fig5}$a$~and~\ref{fig5}$b$.  In
Figures~\ref{fig5}$c$~and~\ref{fig5}$d$, the superimposed red
curve shows angular velocity $\overline{u}_A=\overline{u}/\cos\phi$.

The mean flow acceleration is such that $\overline{u}_A$ evolves
towards $c_A$ \citep[e.g.][]{Andrews87,Holton72}. In
Figure~\ref{fig5}$c$, because $c_A \lesssim \overline{u}_A$
everywhere, the equatorward propagating Rossby waves can only
decelerate the zonal flow. The waves, which are generated at the cores
of the jets, propagate equatorward until they encounter a critical
layer on the equatorward flanks of the jets.  In the critical layer,
Rossby waves break and deposit easterly momentum.  Hence, no
superrotation is present in the control simulation. In contrast, in
Figure~\ref{fig5}$d$ (RTG simulation), the $m\!\approx\! 6$ modes that
propagate poleward from their source region at $\phi \approx \pm
5\degree$ deposit westerly momentum near the equator where $c_A\gtrsim
\overline{u}_A$. These waves break at $\phi \approx \pm
15$--$20\degree$ and decelerate the flow there.  While the
$m\!\approx\! 6$ modes are only able to accelerate the flow at $\phi
\approx \pm 5\degree$, the $m \!\approx\!  2$ modes (with $c_A\approx
30$\,m\,s$^{-1}$) accelerate the zonal flow at the equator by
converging westerly momentum and forcing $\overline{u}_A$ towards
$c_A$ throughout the duration of the simulation.

\begin{figure*}
\centering
\includegraphics[height=5.7cm,width=0.9\textwidth]{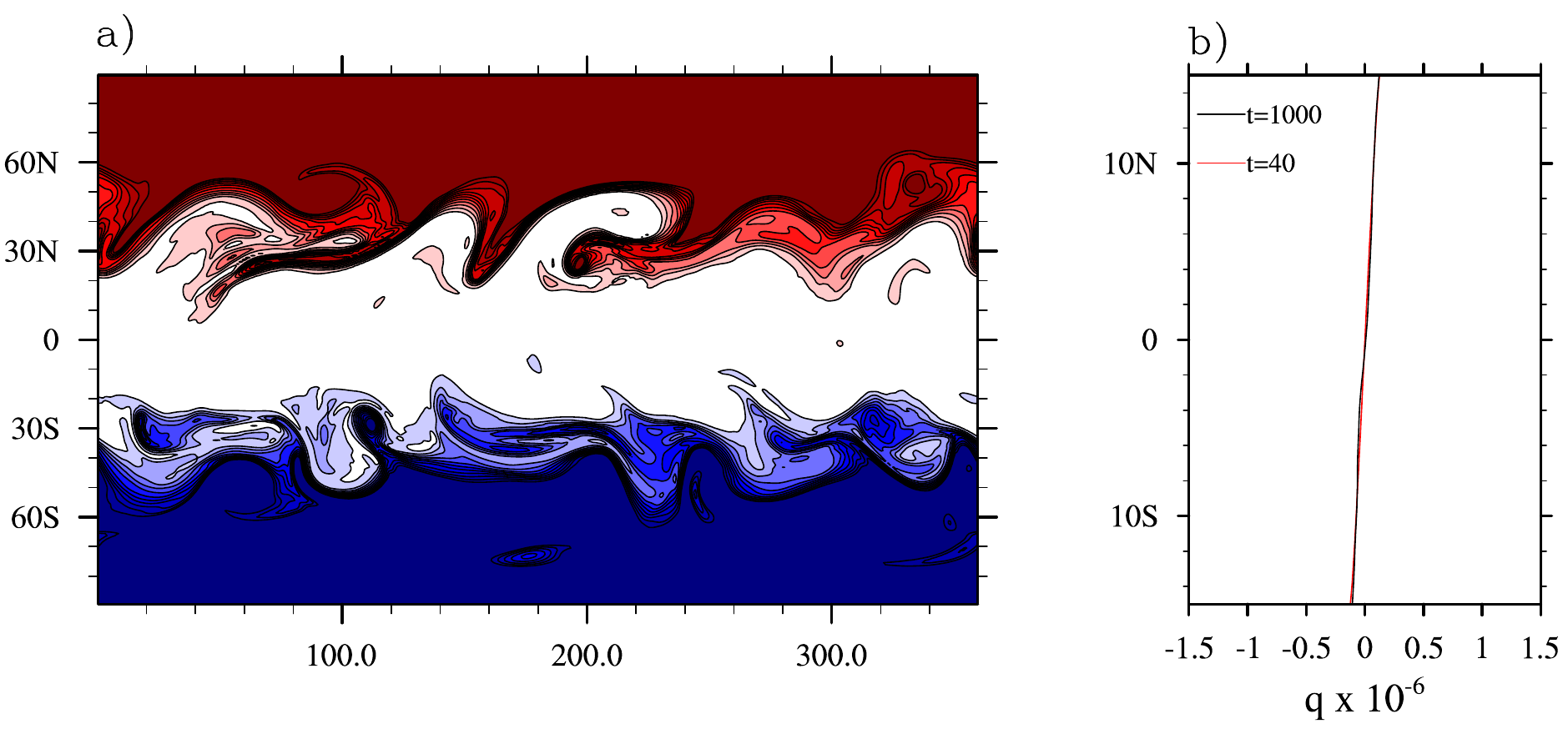}\\
\includegraphics[height=6.3cm,width=0.9\textwidth]{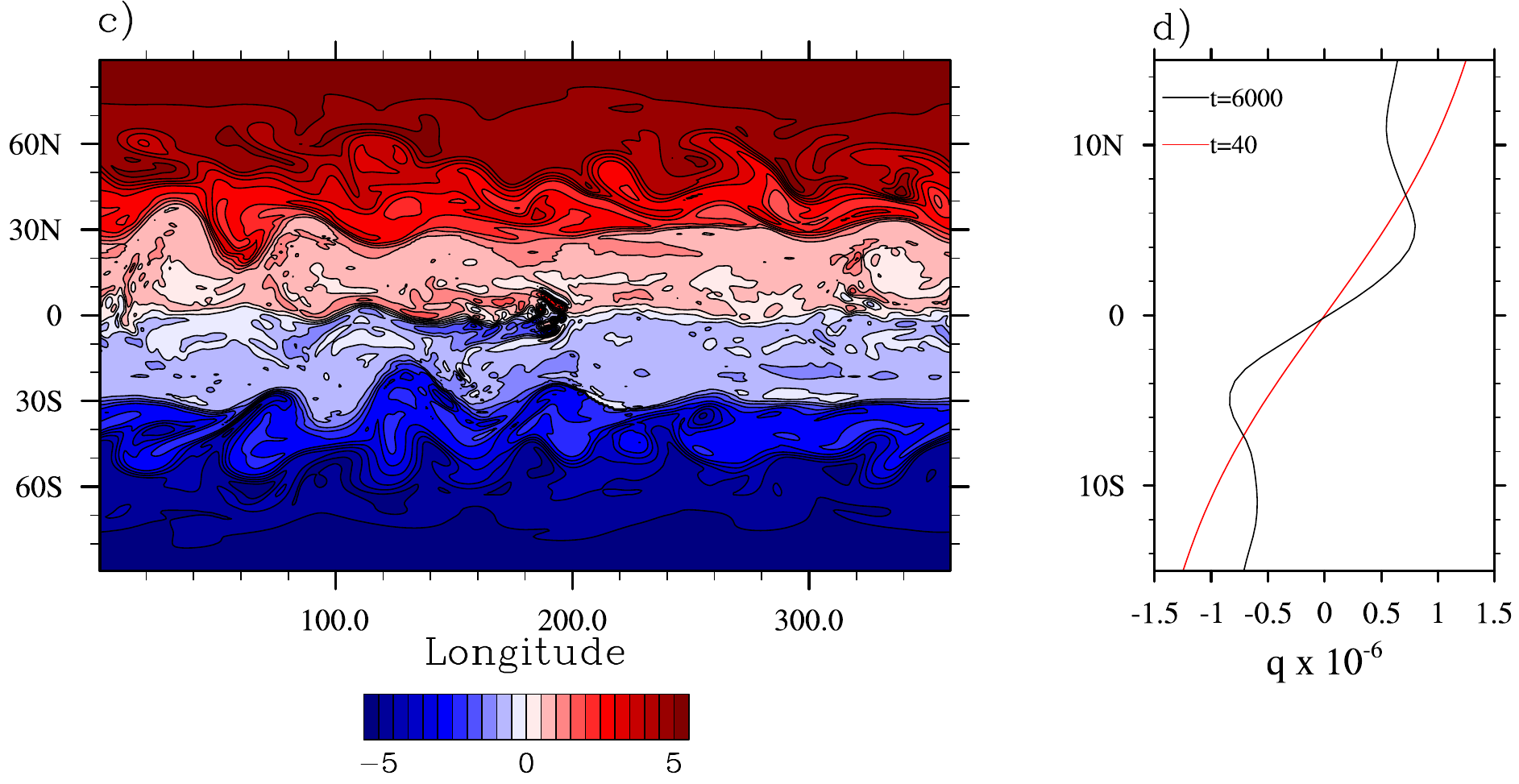}
\caption{Snapshot of Ertel potential vorticity $q \times
  10^{-6}$\ [K\,m$^2$\,kg$^{-1}$\,s$^{-1}$] and its zonal-mean
  $\overline{q}$ for the control simulation ($a$ and $b$) and RTG
  simulation ($c$ and $d$) at T170L30 resolution.  The $q$ fields are
  shown at $t = 1000$\,days in ($a$) and at $t = 6000$\,days in ($c$)
  in cylindrical-equidistant projection on the $\Theta = 315$\,K and
  $\Theta = 300$\,K isentropes, respectively.
  The $m\!  \approx\!  6$ mode can be seen in ($a$) and ($c$), as well
  as the presence of equatorial $q$-gradients and blobs in ($c$)
  (only).  $\overline{q}$ in the equatorial region is shown in ($b$)
  and ($d$) on the same isentropes and at the same times as in ($a$)
  and ($c$).  $\overline{q}$ at $t = 40$\,days (red) is also shown.
  Superrotation is generated in the RTG simulation as a consequence of
  extrusion of high $q$ from higher latitudes to the equator in the
  northern hemisphere via stirring.  In contrast, notice the absence
  of $q$-stirring at the equator in the control simulation.
}\label{fig6}
\end{figure*}
\begin{figure*}
\centering
\includegraphics[width=\textwidth]{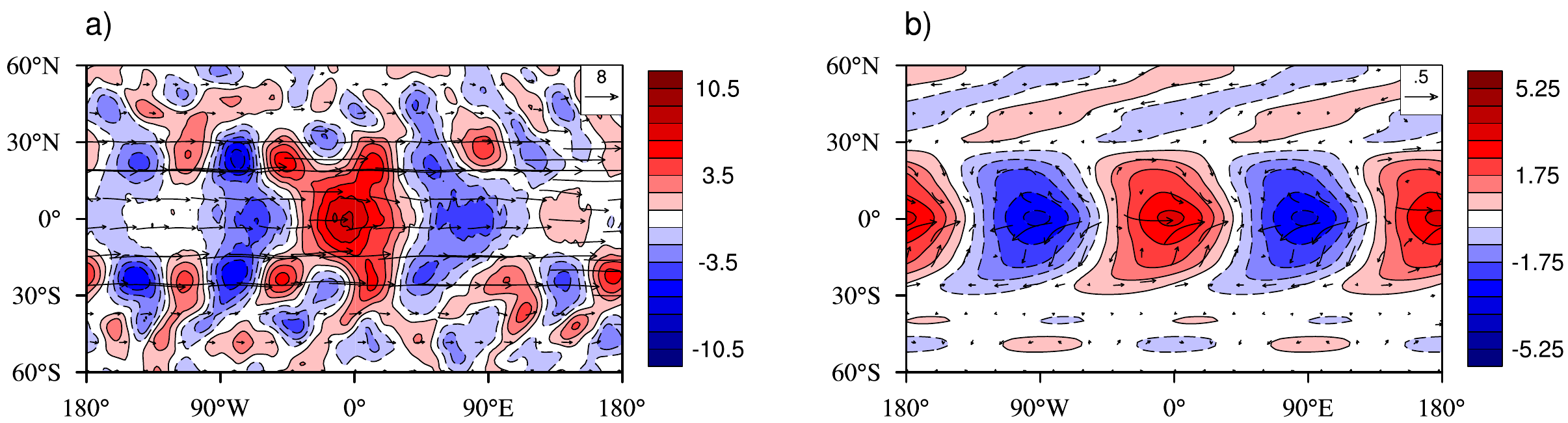}
\caption{Composites of perturbation geopotential height
  $\Phi'$\ [m$^{2}$~s$^{-2}$] with perturbation horizontal wind
  vectors overlaid for the RTG simulation at T170L30 resolution. The
  fields are shown in cylindrical-equidistant projection for
  $|\phi|<\pm 60\degree$ at the $p = 500$\,hPa level.  The unfiltered
  fields are shown in ($a$). The fields filtered so that only $m=2$
  signal remains are shown in ($b$).  Note the resemblance of the
  filtered fields with the classic Kelvin wave solution.
}\label{fig7}
\end{figure*}
\begin{figure*}
\centering $\begin{array}{cccc}
  \includegraphics[width=0.5\textwidth]{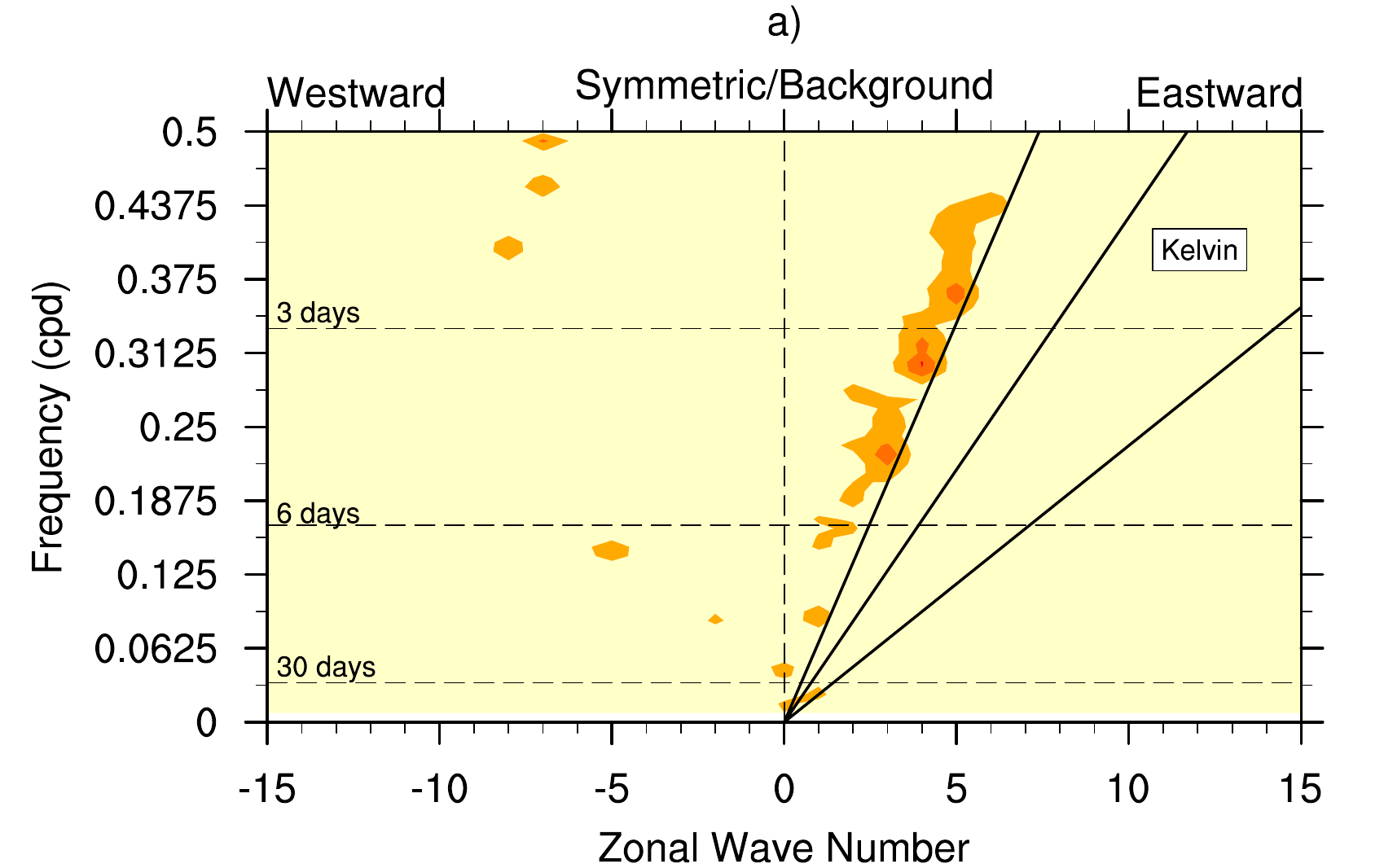}\hspace{0.5cm}
  \includegraphics[height=5.5cm,width=0.5\textwidth]{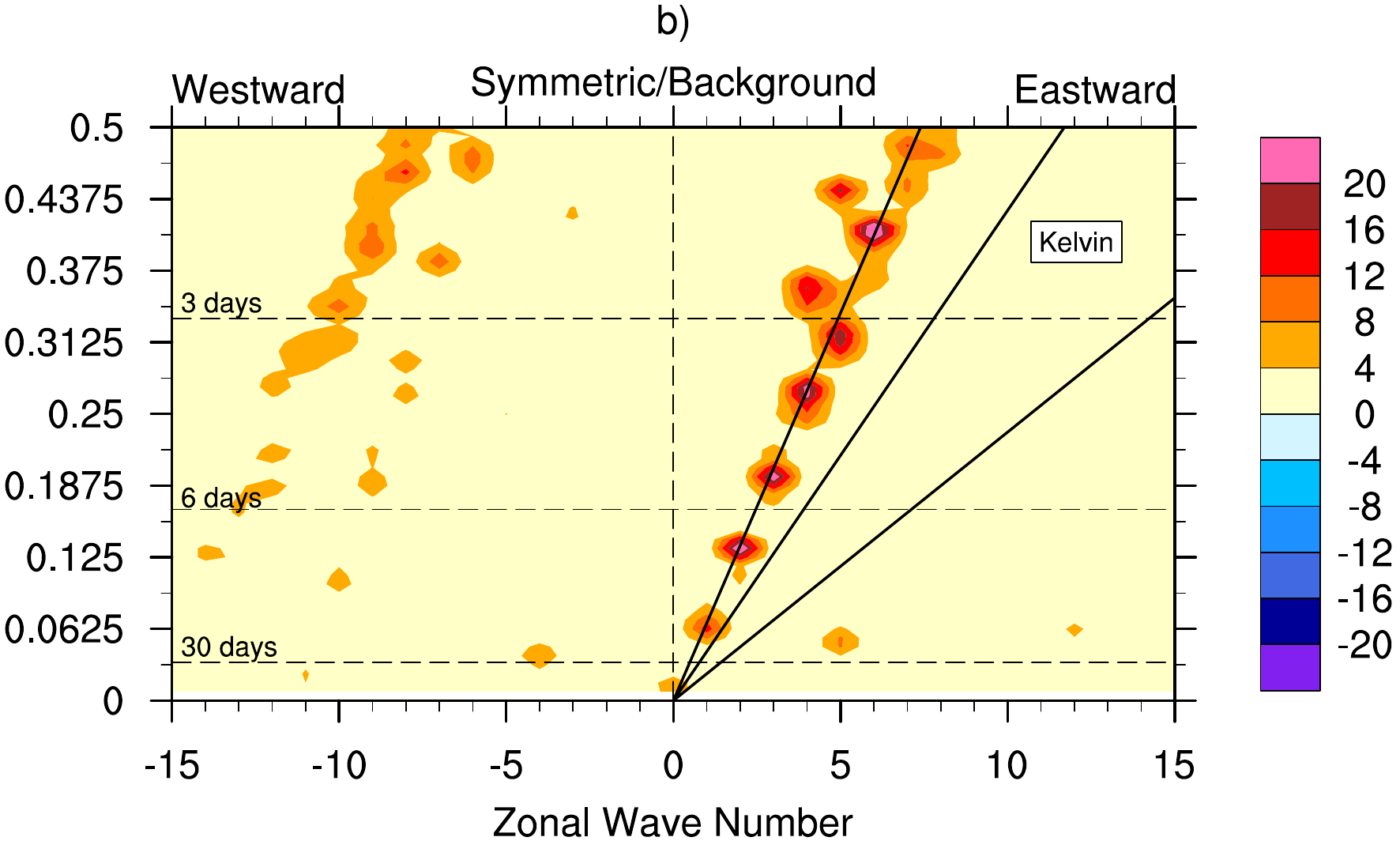}\\ 
  \includegraphics[width=0.5\textwidth]{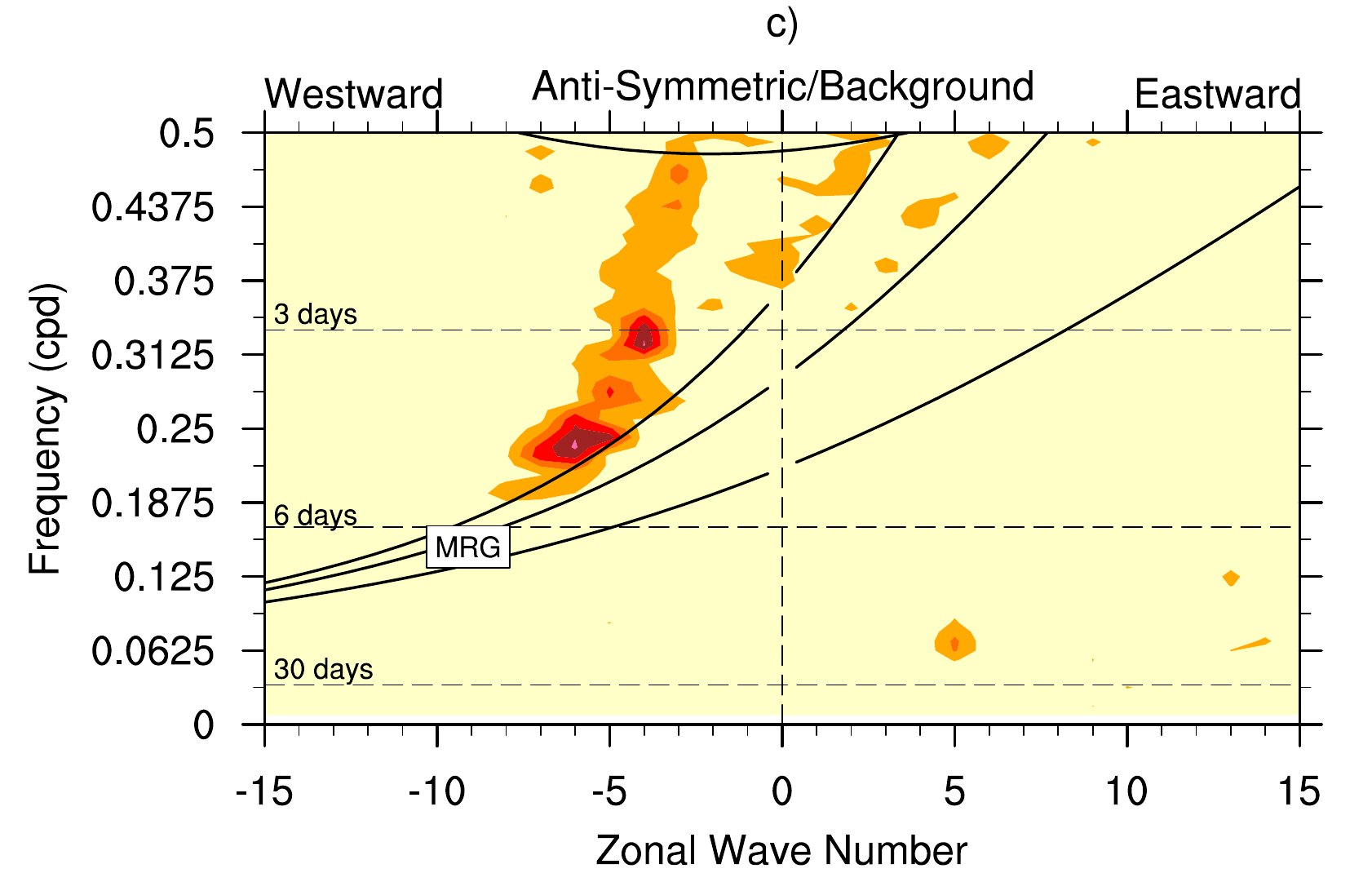}\hspace{0.5cm}
  \includegraphics[height=5.5cm,width=0.5\textwidth]{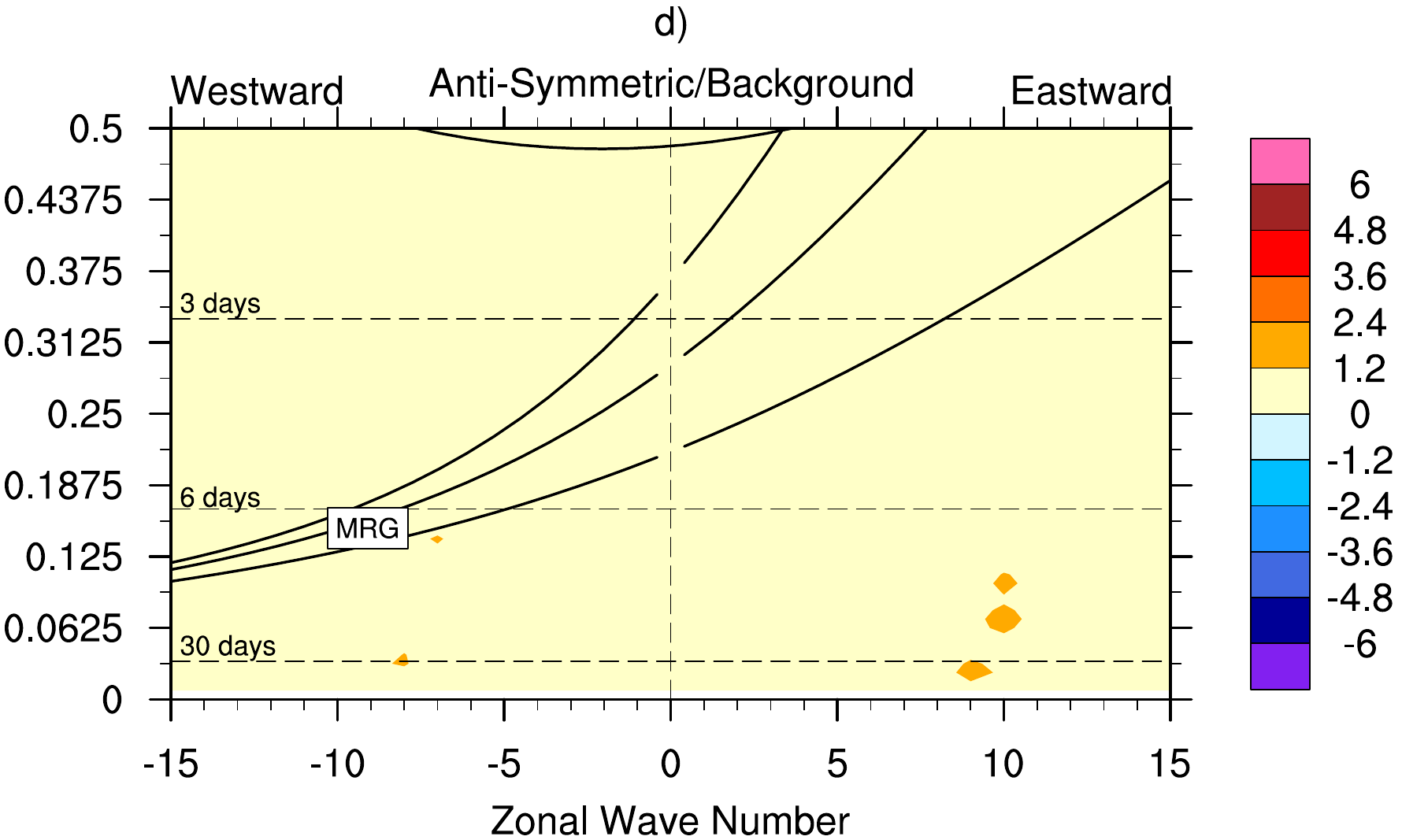}\\
\end{array}$
\caption{Symmetric and antisymmetric mode Wheeler-Kiladis diagrams of
  the geopotential height field in the equatorial region
  ($|\phi|<10\degree$) for the control ($a$\,and\,$c$) and RTG
  ($b$\,and\,$d$) simulations at $p=500$\,hPa. The linear dispersion
  curves for Kelvin and mixed Rossby-gravity waves for equivalent
  depths $h_0 =12, 40,$ and $100$\,m are overlaid on the
  diagrams. Notice the $h_0 = 100$\,m Kelvin wave in the RTG
  simulation.}\label{fig9}\end{figure*}
\begin{figure*}
\centering
\includegraphics[width=\textwidth]{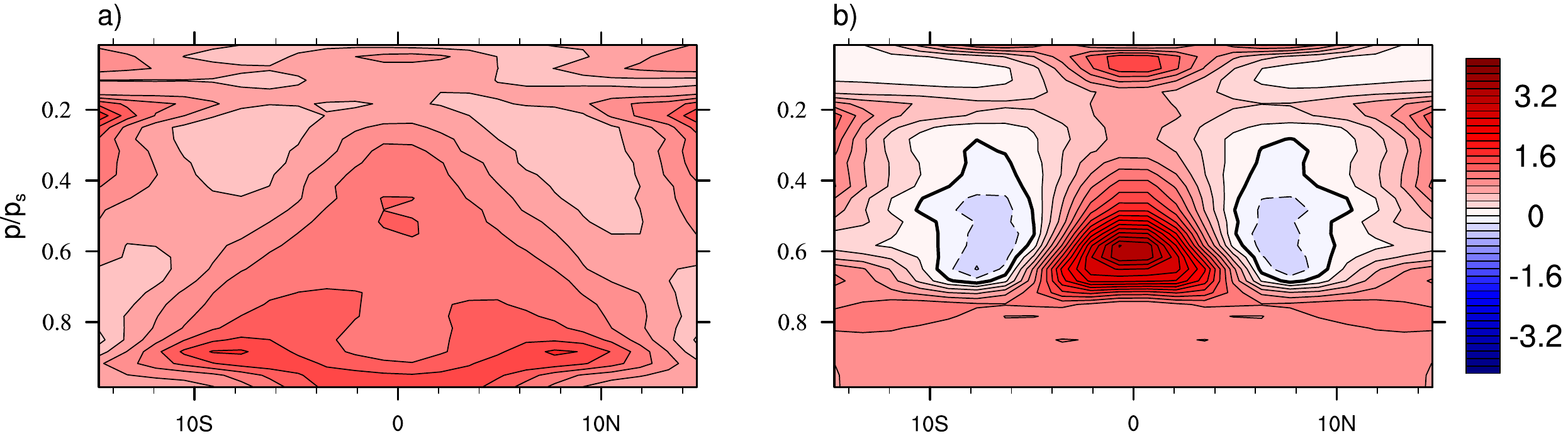}
\caption{Time- and zonal-mean meridional gradient of quasi-geostrophic
  potential vorticity $\del \overline{Q}/\del \phi $ [normalized by $2
    \Omega$], pressure vs. latitude, for the control ($a$) and RTG
  ($b$) simulations at T170L30 resolution in the equatorial
  region. Negative $\del \overline{Q}/\del \phi$ values are dashed
  (and in blue) and positive are solid (and in red). The zero contour
  is drawn with double thickness.  Note the reversal of $\del
  \overline{Q}/\del \phi$ sign in the equatorial region of the RTG
  simulation. }\label{fig8}\end{figure*}
\begin{figure*}
\centering
$\begin{array}{cccc}
  \includegraphics[width=0.66\textwidth]{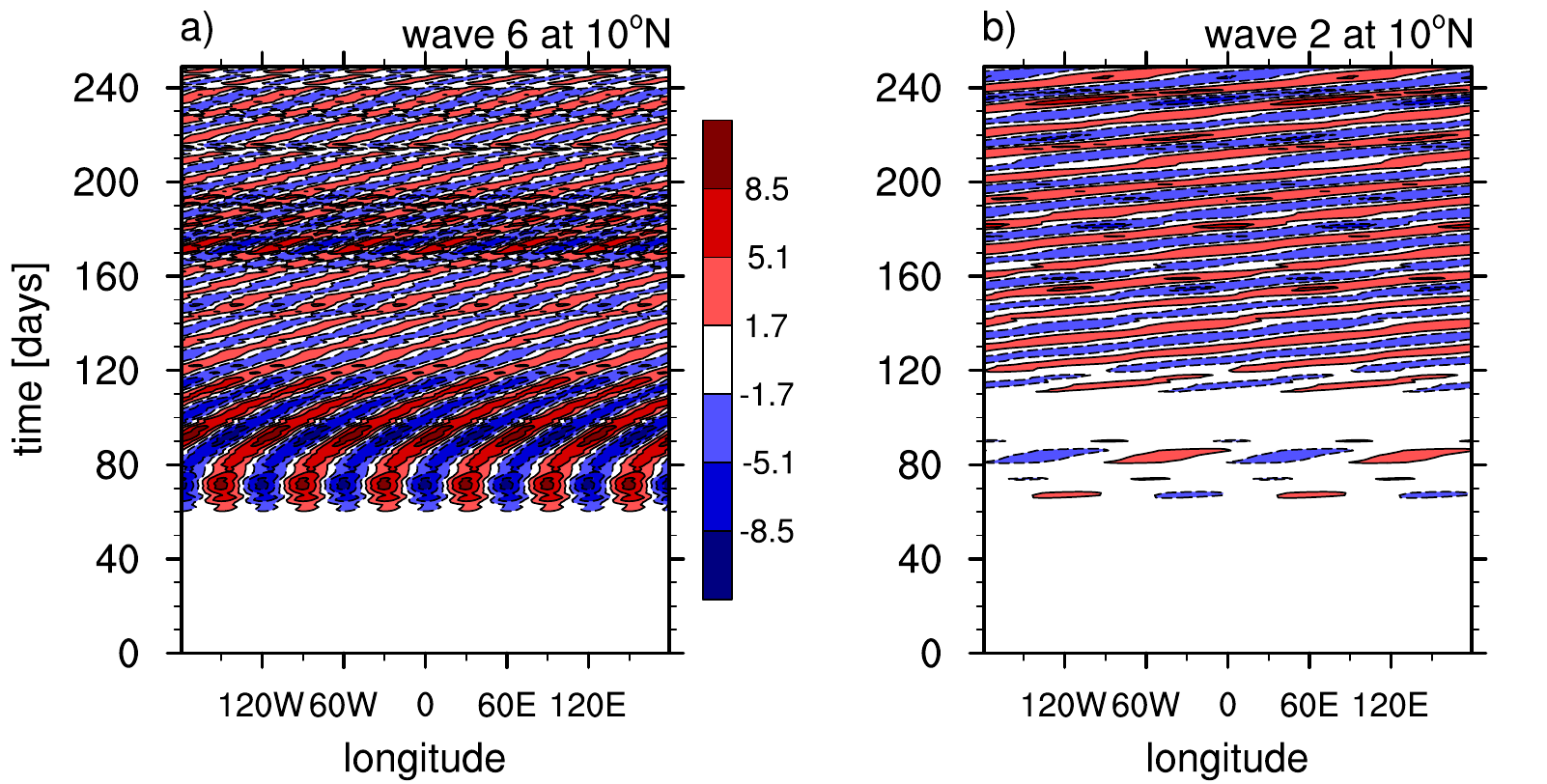}
  \includegraphics[width=0.33\textwidth]{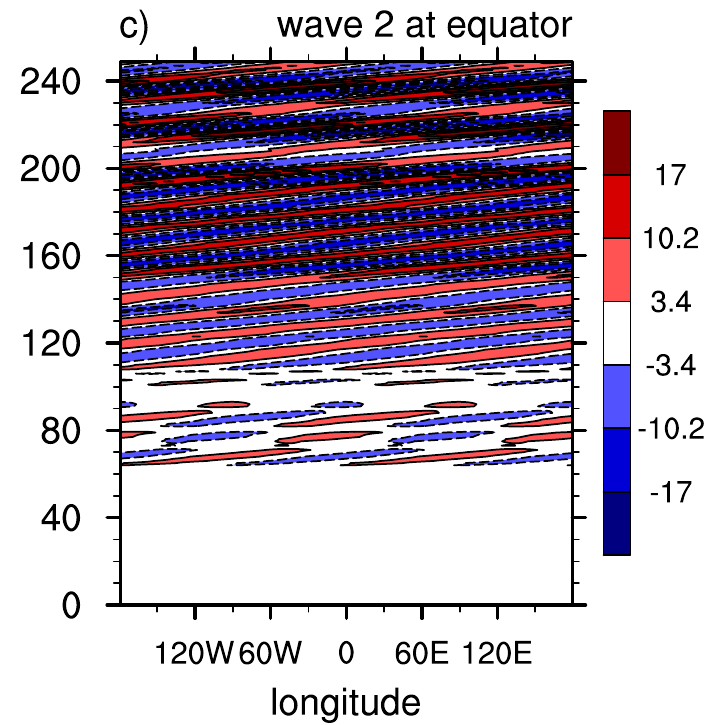}
\end{array}$
\caption{Hovm{\"o}ller diagrams of the perturbation
    geopotential height field $\Phi'$\,[m$^{2}$~s$^{-2}$] at
    $p=500$\,hPa in the beginning of the RTG simulation. The $\Phi'$
    field filtered for $m=6$ is shown at $\phi =10\degree$\,N in
    $a$). The $\Phi'$ field filtered for $m=2$ is shown at $\phi
    =10\degree$\,N in $b$) and at the equator in $c$).  The $m=6$
    subtropical Rossby waves are generated before the equatorial $m=2$
    Kelvin waves.}\label{fig10}\end{figure*}

For a more detailed view of the flow field at equilibrated stage, a
snapshot of Ertel potential vorticity, $q = [(\zeta +
  f)\!\cdot\!\nabla\theta]\,/\,\rho$, is shown in Figure~\ref{fig6};
here $\rho$ is the density and the projection shown is
cylindrical-equidistant centered on the equator.  Figure~\ref{fig6}$a$
shows $q$ for the control simulation at $t = 1000$\,days, on the
$\theta=315$\,K isentrope.  Figure~\ref{fig6}$c$ shows $q$ for the RTG
simulation at $t = 6000$\,days, on the $\theta=300$\,K isentrope. Both
isentropes lie approximately on the $p=500$\,hPa surface in the
tropics.  Both simulations are at T170L30 resolution.  To illustrate
how the PV distribution changes from the initial stage, the zonal-mean
PV ($\overline{q}$) for snapshots in
Figures~\ref{fig6}$a$~and~\ref{fig6}$c$ is shown in the equatorial
region in Figures~\ref{fig6}$b$~and~\ref{fig6}$d$ (black curves)
together with $\overline{q}$ at $t=40$~days (red curve).  Note that at
$t=40$~days the unaveraged flow in both simulations is nearly
axisymmetric, but becomes increasingly non-axisymmetric over time.

The $m\!\approx\! 6$ structure can clearly be seen in
Figures~\ref{fig6}$a$~and~\ref{fig6}$c$, but stirring of $q$ occurs
closer to the equator in the RTG simulation than in the control
simulation (\ref{fig6}$c$~and~\ref{fig6}$a$, respectively).  This is
consistent with the diagnostics in Figure~\ref{fig5} discussed above.
In the control simulation no stirring is taking place \emph{in the
  equatorial region} as the zonal-mean PV changes little from the
zonally-symmetric PV at $t=40$~days (see
Figure~\ref{fig6}$b$). Crucially, in the RTG simulation, stirring
results in a band of steep $q$-gradient, which is pushed to the
equator; a corresponding band is not present in the control
simulation. The band is not zonally-symmetric and often appears with a
blob of $q$ that appears to have broken off from large-amplitude
undulations of the high-gradient band to the north (south) of the
equator in the northern (southern) hemisphere. In addition to inducing
a `$q$-jump' at the equator, the undulations and concentrations of
large-amplitude PV at the equator, act as sources of enhanced
superrotation. The blobs of $q$ at the equator translate
longitudinally faster than the zonal mean zonal wind and manifest
themselves in the convergence spectrum as the $m \approx 2$ Kelvin
wave-like features (see Figure~\ref{fig5}$b$). Note
  that in the presence of meridional shear in the background flow,
  Kelvin waves perturb meridional velocity and hence $q$
  \citep[e.g.][]{Iga04}.

While the $m\approx 2$ features lack a coherent wave-like structure in
the $q$ field in Figure~\ref{fig6}$c$, the wave-like behaviour is more
clearly visible in the perturbation geopotential height field $\Phi'$
in the equatorial region of the RTG simulation. Figure~\ref{fig7}
shows composites of $\Phi'$ and perturbation horizontal wind vectors
on the $p=500$~hPa surface for the RTG simulation. The composites are
constructed from the $m-\varpi$ spectrum from the equilibrated
fields. The spectrum is filtered by removing all westward propagating
signal, eastward propagating signals with frequencies below 1/30
day$^{-1}$ and above 1/4 day$^{-1}$, and $m$ below 1 and above 6. An
inverse transform is then applied to reconstruct the filtered time
series and the entire fields are regressed against the filtered time
series at a base point on the equator, where the wave variance is
maximum. In ($a$) the complete composited fields are shown and in
($b$) the fields have been filtered to accentuate the $m=2$ signal.

The filtered fields in particular, resemble the classic Kelvin wave
solution for the equatorial beta-plane shallow water model
\citep[e.g.,][]{Matsuno66}. By examining the time-lagged composites,
we have found that the $m=2$ signal in Figure~\ref{fig7}$b$ is only
coherent up to $\phi \approx \pm 15\degree$ (not shown). Linear
resonance between equatorial Kelvin waves and subtropical Rossby waves
via ageostrophic instability is proposed by \cite{Wang14} to drive
superrotation under zonally-symmetric thermal forcing. Such
instability would have a distinct Kelvin-wave like structure at the
equator (see e.g. Figure~1 in \cite{Wang14}).  However, the linear
Rossby-Kelvin instability also requires matching phase speeds between
the two types of waves, since the wavenumber $m$ is fixed in the
linear resonance analysis. In Figure~\ref{fig5}$d$, however, the phase
speed of the subtropical Rossby waves ($c_A\approx10$\,m\,s$^{-1}$) is
clearly different than the phase speed of the equatorial Kelvin waves
($c_A\approx30$\,m\,s$^{-1}$). This is so even during
  the acceleration stage of the well-resolved RTG simulation (not
  shown). Hence, the linear Rossby-Kelvin wave
instability is not the likely mechanism generating and maintaining
superrotation in the RTG simulations. Instead, as in
\citep{Williams03}, it appears that here the
  barotropic instability on the equatorward flank of the westerly
  subtropical jets stirs the flow and excites equatorial Kelvin
waves.

Kelvin waves in the RTG simulation can be further identified by Fourier
transforming equatorial geopotential height field $\Phi$ in space and
time and depicting the resulting spectra in the Wheeler-Kiladis
diagrams as a function of $\varpi$ and $m$ \citep{wheeler1999}.  Here
ten 96-day time windows are taken to obtain the $\Phi$ spectrum.  In
Figure~\ref{fig9}, both symmetric ($a$~and~$b$) and antisymmetric
($c$~and~$d$) Wheeler-Kiladis diagrams for the control ($a$~and~$c$)
and RTG simulations ($b$~and~$d$) at T170L30 resolution are shown at
the $p=500$\,hPa level.  Overlaid on the symmetric diagrams are the
linear dispersion curves for the Kelvin waves and on the antisymmetric
diagrams the linear dispersion curves for the mixed Rossby-gravity
waves for equivalent depths of $h_0 =12,40,$ and $100$~m,
respectively.

For the control simulation, only a weak Kelvin wave signal is present
at the equator (Figure~\ref{fig9}$a$) and the spectrum is dominated
by the westward propagating mixed Rossby-gravity waves with $h_0 =
100$\,m (Figure~\ref{fig9}$c$). For the RTG simulation, a clear $h_0
=100$\,m Kelvin wave signal dominates the spectrum
(Figure~\ref{fig9}$b$) with little or no westward propagating wave
activity. The phase speed of the $h_0 =100$\,m Kelvin waves is $\approx\!
30$\,m\,s$^{-1}$, suggesting that Kelvin waves are important in
driving and maintaining superrotation in the RTG simulations
(cf. Figure~\ref{fig5}$d$). We note that the equatorial $\Phi$ spectra
is of lower amplitude in simulations with lower horizontal resolution
after the acceleration stage.

To establish the role of shear instabilities in Kelvin wave
excitation, it is instructive to examine the meridional gradient of
the zonal-mean PV in the equatorial region. As in \cite{Williams03},
we show quasi-geostrophic time- and zonal-mean PV $\del
\overline{Q}/\del \phi$ (scaled by 2$\Omega/a$) in
Figure~\ref{fig8} for the control ($a$) and RTG ($b$) simulations,
respectively.  In Figure~\ref{fig8}$b$ $\del \overline{Q}/\del
\phi$ changes sign on the equatorward flank of the subtropical jet, at
$\phi = \pm 5 \degree$ in the $p = [200, 700]$\,hPa vertical region
throughout the RTG simulation. The reversal of the gradient at $\phi =
\pm 5 \degree$ is due to the barotropic component (i.e. the meridional
gradient of the absolute vorticity) of $\del \overline{Q}/\del \phi$
(not shown).  Hence, the source of the $m \!\approx\!  6$ Rossby waves
on the equatorward flanks of the jets and the equatorial $m
\!\approx\!  2$ Kelvin waves is likely the barotropic, Rayleigh-Kuo,
instability. The (latitudinal) deformation length scale $L_{\text{D}}$
of an equatorially trapped Kelvin wave is roughly $L_{\text{D}} =
\sqrt{c/\beta}$, where $\beta = 2 \Omega/ a$.  For the wave with $c
\approx 30$\,m\,s$^{-1}$
(cf. Figures~\ref{fig5}$d$~and~\ref{fig9}$b$), $L_{\text{D}} \approx
1200$\,km or $\phi \approx \pm 10\degree$.  Therefore, to excite
equatorially trapped Kelvin waves, stirring must occur within the
$\phi \approx \pm 10\degree$ band centered on the equator. This occurs
in the RTG simulation.  For the control simulation, the necessary
condition for neither the barotropic (see Figure~\ref{fig8}$a$) nor
the baroclinic instability is satisfied in this region in the time-
and zonal-mean, which likely explains the weak equatorial Kelvin wave
signal in the control simulation (see Figure~\ref{fig9}$a$).

That stirring in the subtropics (within $L_{\text{D}}$) by the
$m\approx 6$ Rossby waves precedes the generation of equatorial
$m\approx 2$ Kelvin waves (with $c\approx 30$\,m\,s$^{-1}$) in the RTG
simulation can be seen in Hovm{\"o}ller diagrams in
Figure~\ref{fig10}. The figure shows $\Phi'$ at $p = 500$\,hPa in the
beginning of the simulation. The $\Phi'$ field filtered for $m=6$ is
shown at $\phi=10\degree$N in ($a$). The $\Phi'$ field filtered for
$m=2$ is shown at $\phi=10\degree$N in ($b$) and at the equator in
($c$). The Kelvin waves are excited at $t\approx 110$\,days
(Figures~\ref{fig10}$b$~and~$c$), after the development of the $m=6$
Rossby waves at $t\approx 80$\,days (Figure~\ref{fig10}$a$). These
waves continue to propagate eastward throughout the simulation (not
shown).

In summary, the following mechanisms contribute to
  generating and/or maintaining superrotation in \emph{well-resolved}
  RTG simulations:
\begin{itemize}
\item{Mechanical stirring inside the latitudinal range of the
  deformation length scale $L_{\text{ D}}$ from the equator, is
  required to excite $m\approx 2$ Kelvin waves under statically stable
  zonally-symmetric thermal forcing\footnote{The
      destabilization of the $m\approx 2$ waves is dependent on the
      background flow (e.g. \cite{Iga04})}.  Here the source of the
  stirring is barotropic instability (with predominantly $m\approx 6$
  signal) on the equatorward flank of the subtropical westerly jet.}
\item{Since Kelvin waves are not excited if the stirring occurs at
  latitudes outside $L_{\text{ D}}$, the specified equilibrium
  temperature profile has to produce and maintain westerly jets that
  are unstable to shear instabilities within $L_{\text{ D}}$ from the
  equator.}
\end{itemize}

\subsection{Resolution and dissipation}\label{sec3.3}

We have seen that superrotation may be generated and maintained by
wave-mean flow interaction.  However, in the RTG simulations, a
stronger superrotation is also clearly produced at lower horizontal
resolution.  This indicates that part of the superrotation is of
numerical origin.  In general, the strength of superrotation depends
sensitively on the horizontal resolution and dissipation.  We have
found that relatively high horizontal resolution (\,$\gtrsim\!$~T170)
is required for numerical convergence -- particularly for the RTG
case.  Among other things, the high resolution ensures accurate
representation of the eddy fluxes.

In this study, simulations are defined to be {\it converged} if both
of the following criteria on the total (column-averaged) equatorial
zonal wind $\langle u_{\rm eq} \rangle$---averaged over the $\phi=\pm
5\degree$ band centered on the equator---are met:
\begin{description}

\item {\it i}) At equilibration, $\langle u_{\rm eq} \rangle$ is
  statistically the same at least at two different horizontal
  resolutions, with a given viscosity coefficient $\nu$.

\item {\it ii}) At equilibration, $\langle u_{\rm eq} \rangle$ is not
  sensitive to the choice of viscosity coefficient $\nu$, at a given
  horizontal resolution.

\end{description}

Criterion~{\it i}) tests for `numerical' convergence while
criterion~{\it ii}) tests for `physical' convergence.  At very high
Reynolds number, a simulation (particularly the large scales in it)
should not be sensitive to the choice of numerical viscosity
coefficient $\nu$, whose primary purpose is to remove enstrophy near
the grid-scale and to prevent the simulation from blowing up.  For
criterion~{\it ii}), $\nu$ values are heuristically chosen from a
`credible' range, by which reference is made to those values that
permit the vorticity field to be neither under-dissipated
(i.e. inundated with grid-scale noise) nor over-dissipated
(i.e. devoid of any strong coherent structures, such as vortices and
sharp gradients).  Experience has shown that a certain amount of
tuning is always necessary and that there always exists a finite range
of values that is reasonably free from gross subjectivity and
satisfies the credibility condition on the vorticity
field.  Each new problem and setup necessitates a thorough
characterization of accuracy and convergence for confidence in the
obtained results.
\begin{figure*}
\centering
   \includegraphics[width=\textwidth]{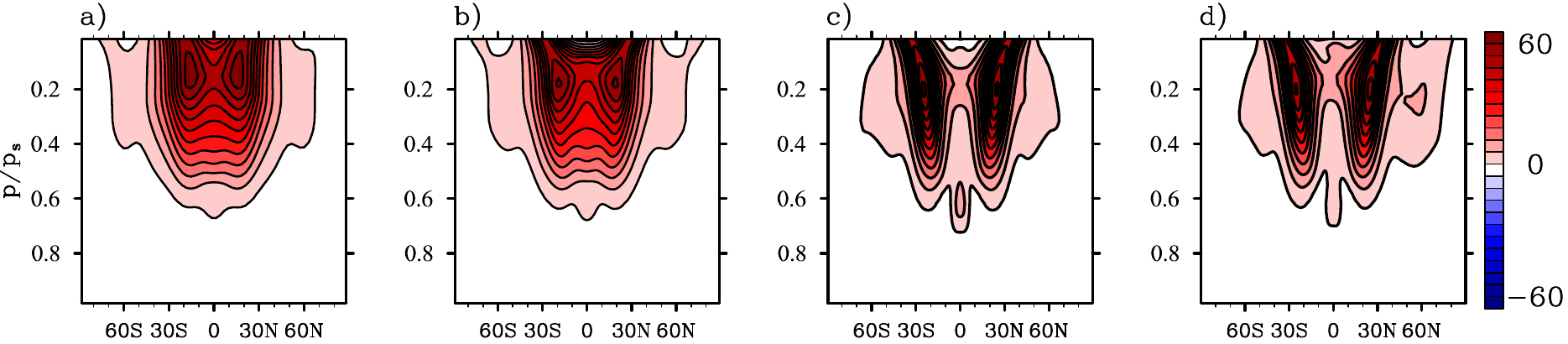}
   \caption{The time- and zonal-mean zonal wind $\overline{u}^*$
     [m\,s$^{-1}$] for the RTG simulation at T42L30 ($a$), T85L30
     ($b$), T170L30 ($c$) and T341L30 ($d$) resolutions.  Contour
     intervals are the same as in Figure~\ref{fig2add}.  Subtropical
     jets diffuse together at the equator much more strongly at low
     resolution.}\label{fig11}
\end{figure*}

Before we discuss $\langle u_{\rm eq} \rangle$, the equilibrated
$\overline{u}^*$ from the RTG simulations at T42L30~($a$),
T85L30~($b$), T170L30~($c$) and T341L30 ($d$) resolutions is presented
in Figure~\ref{fig11}.  The figure clearly shows that RTG simulations
are not converged at resolutions lower than T170L30, as they do not
satisfy criterion~{\it i}).  The T170 and T341 results are very
similar.  In contrast, the equilibrated $\overline{u}^*$ for the
control simulations can exhibit qualitatively the same behaviour at
the same four horizontal resolutions presented (although they are
still not converged, as is also discussed below).  Interestingly, as
can be seen in the figure, superrotation is considerably stronger at
lower horizontal resolution in the RTG simulations, as already
mentioned.

\begin{figure*}
\centering
 $\begin{array}{cc}
    \includegraphics[height=5cm,width=0.5\textwidth]{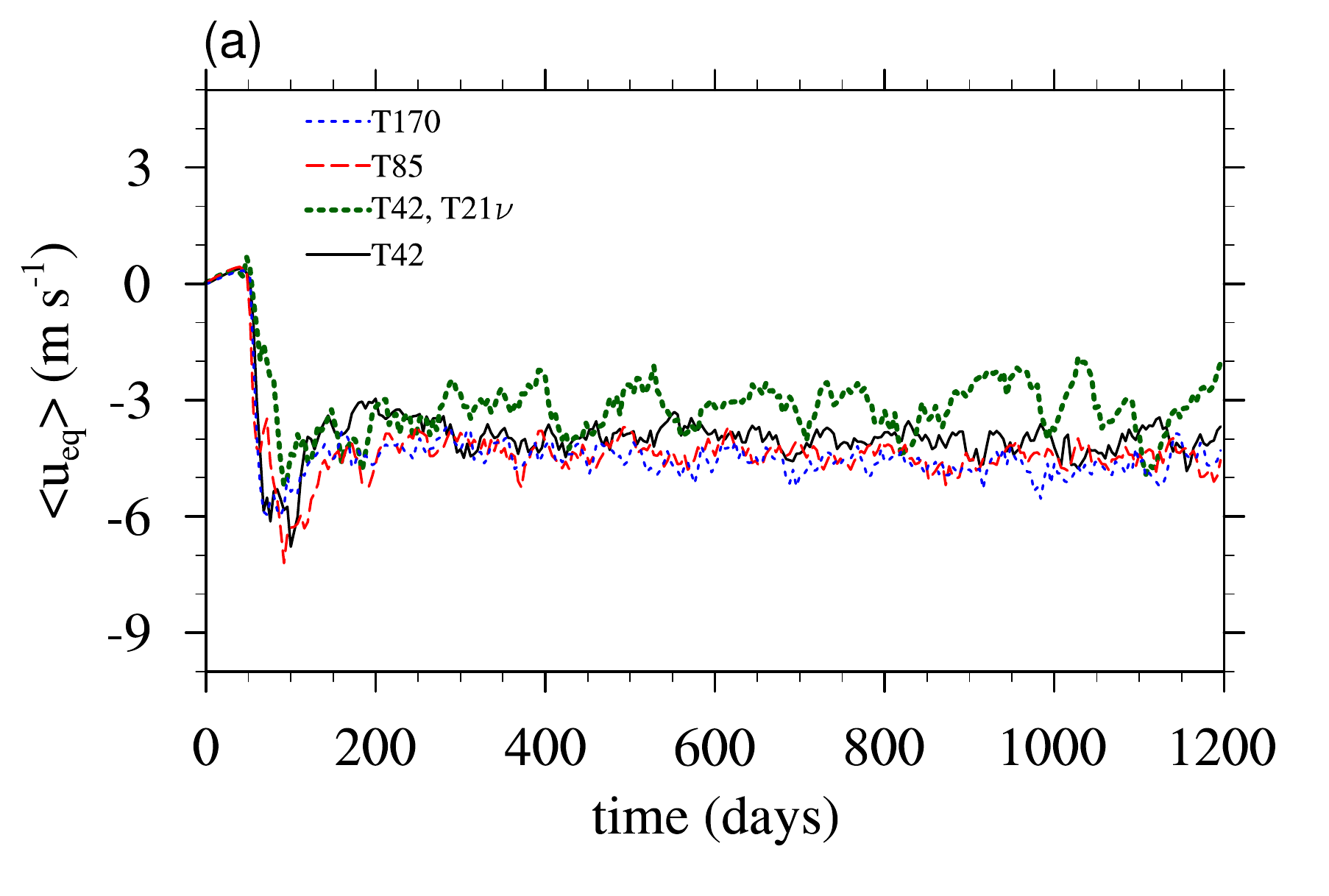}
   \includegraphics[height=5cm,width=0.5\textwidth]{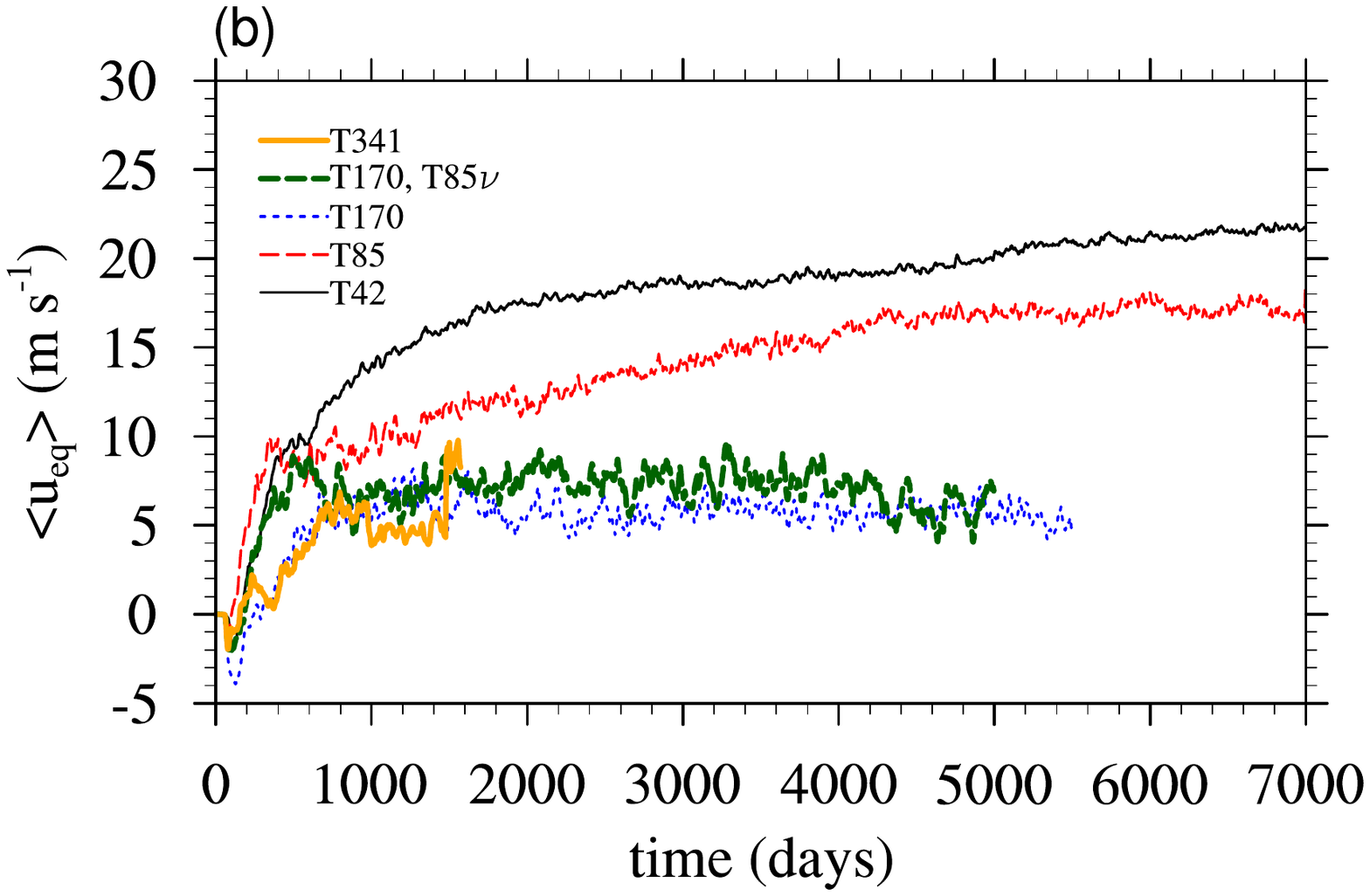}
  \end{array}$\\
  \caption{Time series of column-averaged zonal wind at the equator
    $\langle u_{\rm eq}\rangle$~[m\,s$^{-1}$] at varying horizontal
    resolutions for the control ($a$) and RTG ($b$) simulations:
    $\langle u_{\rm eq}\rangle$ for \{T42L30, T85L30, T170L30,
    T341L30\}~resolutions are shown in \{black, red, blue, yellow\},
    respectively.  The green curve in $a$) shows the control
    simulation performed at T42 horizontal resolution with $\nu$ from
    a T21 resolution simulation, which is 25 times larger than in the
    `nominal' T42 simulation (black curve).  The green curve in $b$)
    shows the RTG simulation performed at T170 horizontal resolution
    with $\nu$ the same as in the T85 resolution simulation (red
    curve).  The magnitude of $\langle u_{\rm eq}\rangle$ is
    insensitive to the horizontal resolution in the control
    simulations -- {\it when properly tuned}, whereas $\langle u_{\rm
      eq}\rangle$ decreases with increasing resolution in the RTG
    simulations.  In the latter simulations, T170L30 simulations with
    different $\nu$ produce nearly-identical results.
  } \label{fig12}\end{figure*}

Figure~\ref{fig12} now shows the time series of $\langle u_{\rm
  eq}\rangle$ from the control (Figure~\ref{fig12}$a$) and
RTG~(Figure~\ref{fig12}$b$) simulations.  For reference, the time
series from the T341L30 RTG simulation is also shown in
Figure~\ref{fig12}$b$.  Note that, with exceptions as noted in the
discussion below, the value of $\nu$ for all the simulations in the
figure is chosen so that the $e$-folding time for the truncation scale
is nominally $\{0.1,\,0.01,\,0.001,\,0.0001\}$\,days at
\{T42,\,T85,\,T170,\,T341\} resolutions, respectively.  The chosen
times correspond to $\nu =$ \{3.0, 0.1, 0.0044, 0.00017\}$\times
10^{37}\!$~m$^8$\,s$^{-1}$, respectively. All simulations are with L30
vertical resolution.

Figure~\ref{fig12}$a$ shows that the convergence criterion~{\it i}) is
fulfilled for the control simulation at T42 resolution.  However, the
control simulation does not fulfill criterion~{\it ii}) at resolutions
lower than T85.  For example, a T42 resolution simulation employing
the $\nu$ value from the T21 resolution simulation (i.e. 25 times
greater value) leads to a significantly different time series
behaviour (cf. black and green lines); the behaviour does not persist
into T85 resolution simulations, as is shown below.  Convergence
criterion~{\it i}) is also fulfilled by the RTG simulation -- but at
much higher (T170) resolution.  Moreover, the value of $\langle u_{\rm
  eq}\rangle$ at equilibration for the converged RTG simulations is
much lower than for the un-converged, lower-resolution simulations, by
as much as 17\,m\,s$^{-1}$ (cf.  black and blue curves in
Figure~\ref{fig12}$b$); this is consistent with the behaviour in
Figure~\ref{fig11}.  In addition, RTG simulations at T42 and T85
resolutions take a considerably longer time to reach equilibration
than that at T170 resolution; in the latter case, equilibration is
reached at $t\!\approx\!1000$\,days, compared with $t\!\approx\!7000$\,days
for the former cases.

\begin{figure*}
\centering
 $\begin{array}{cc}
    \includegraphics[height=5cm,width=0.5\textwidth]{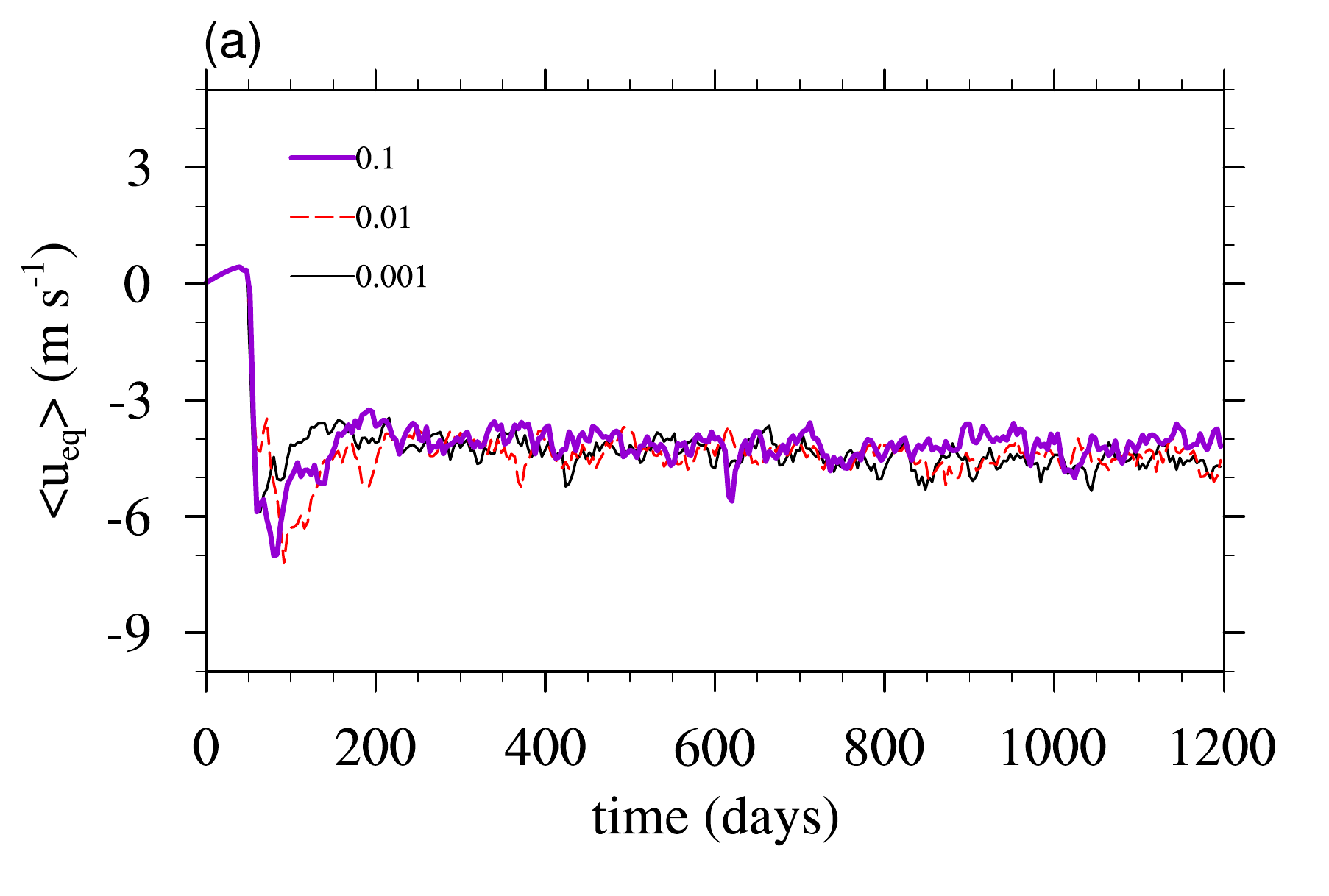}
   \includegraphics[height=5cm,width=0.5\textwidth]{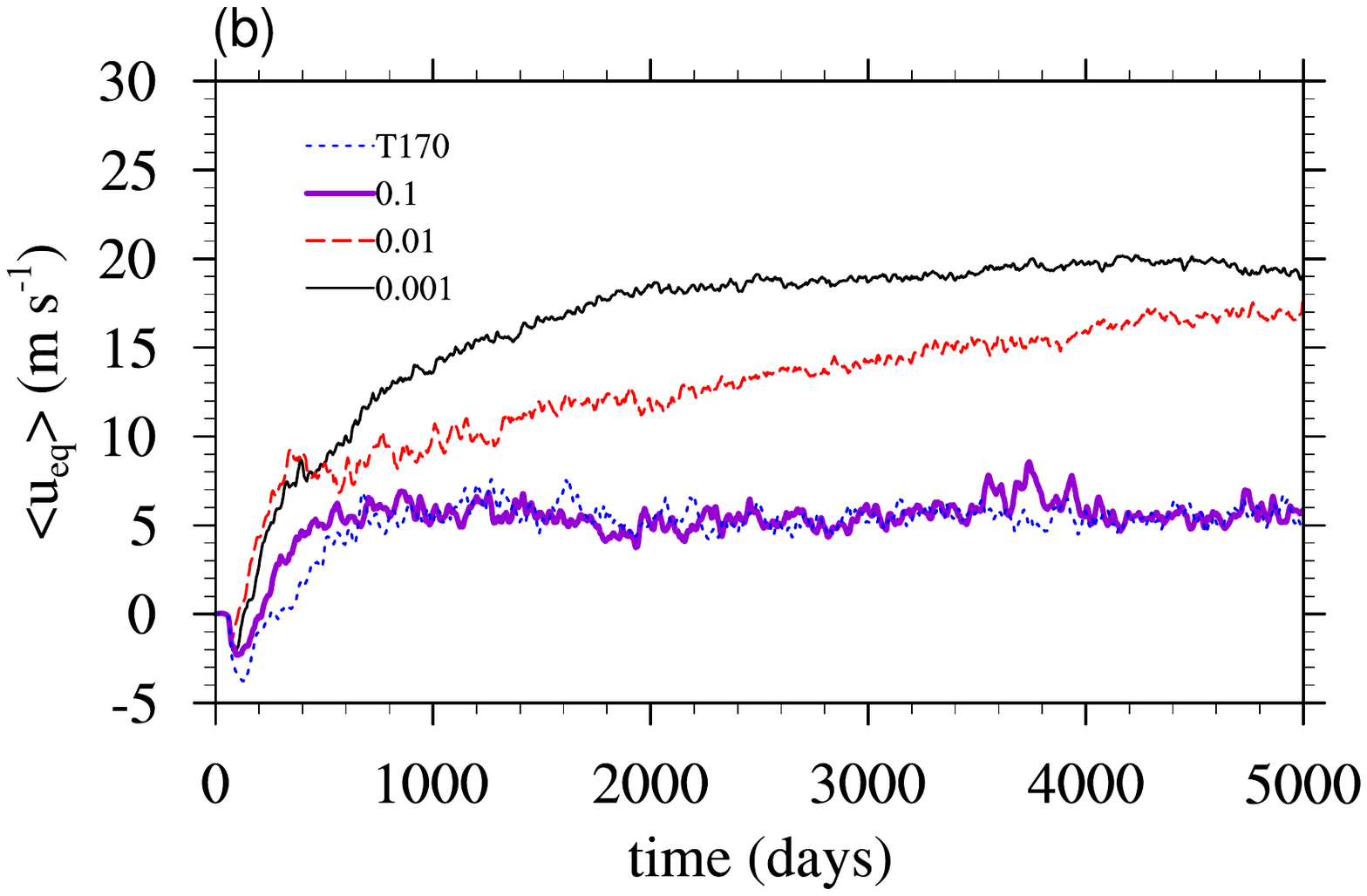}
  \end{array}$\\
  \caption{Time series of $\langle u_{\rm eq} \rangle$~[m~s$^{-1}$] at
    T85L30 resolution with varying $\nu$ for the control ($a$) and RTG
    ($b$) simulations.  The viscosity coefficient $\nu$ is chosen so
    that $\tau_{\rm d}$ is 0.001~days (black curve), 0.01~days (red
    curve) and, 0.1~days (purple curve).  In addition, $\langle u_{\rm
      eq}\rangle$ from the nominal T170L30 simulation is plotted in
    $b$).  Note the insensitivity of the control simulations to $\nu$
    but the reduction of $\langle u_{\rm eq}\rangle$ with decreasing
    $\nu$ in the RTG simulations.  Note also that T170L30 results are
    the same as T85L30 results with $\tau_{\rm d} = 0.1$\,days in
    $b$).  }\label{fig13}\end{figure*}
Throughout this work, a larger value of $\nu$ is used when performing
a given simulation at a lower horizontal resolution -- as is commonly
practiced in numerical simulations. Alternatively, a smaller value of
$\nu$ is used when performing a given simulation at a higher
horizontal resolution.  The practice is primarily carried out for
numerical stability and/or `spectrum tuning' reasons, and it results
in different damping times for a given wavenumber at the different
resolutions.  However, as advocated in \cite{Polvani04} and
\cite{Polichtchouk12}, the same value of $\nu$ should be used for all
resolutions in order to `cleanly' assess numerical convergence: each
wavenumber, up to the truncation, then experiences the same
dissipation rate for a given simulation.  This philosophy is closely
related to convergence criteria~{\it i})~and~{\it ii}). All imply that
once simulation is adequately resolved, including additional, higher
wavenumbers in the simulation would not significantly affect the
large-scale flow. Criterion~{\it ii}), in particular, ensures that
most `stable' value of $\nu$ is used when assessing convergence, since
there exists a finite range of values which prevent a simulation from
blowing up. Of course, all this assumes weak (or absence of) spectral
non-locality of the flow, which generally appears not to be the case
in practice.

Consider the green and red curves in Figure~\ref{fig12}$b$, showing
two RTG simulations at different horizontal resolutions (T85 and T170)
using the same $\nu$ value.  It is clear that the RTG simulation is
not converged at T85 resolution since the two curves deviate from each
other; on the other hand, the curves for T170 and T341 resolutions,
subjected to the same test, do match very closely. In fact, in
general, the disagreement is much stronger, than in the control case
(cf. green and black curves in Figure~\ref{fig12}$a$).  Clearly, the
`extra modes' included in the T170 resolution affect $\langle u_{\rm
  eq}\rangle$ quantitatively in the RTG simulations.

Figure~\ref{fig13} illustrates the above points more explicitly with
simulations at T85L30 resolution, for the control
(Figure~\ref{fig13}$a$) and RTG (Figure~\ref{fig13}$b$) cases.
Criterion~{\it i}) is met for the control case at T85 resolution, and
has already been discussed above.  Figure~\ref{fig13}$a$ demonstrates
that criterion~{\it ii}) is also met: the simulations are not
sensitive to variations in $\nu$, suggesting the control case is
converged at T85 resolution.  In contrast, Figure~\ref{fig13}$b$
demonstrates that criterion~{\it ii}) is not met for the RTG case: the
simulations are acutely sensitive to variations in $\nu$.  Moreover,
superimposing the time series from the T170 resolution simulation
illustrates that it is possible to erroneously conclude that
criterion~{\it i}) is met (see the blue and purple curves), if $\nu$
were not varied at the T85 resolution; and significantly, the
equatorial superrotation is surprisingly weaker at the higher
resolution.  It is important to reiterate at this point that the
`correct' $\nu$ at T85 resolution cannot be determined a priori: it is
only after performing the RTG simulation at T170L30 resolution and
establishing that the converged simulations have $\langle u_{\rm
  eq}\rangle \approx 5$\,m\,s$^{-1}$ that it is possible to tune $\nu$
for an `optimal' simulation.

\begin{figure*} \centering
$\begin{array}{cc}
    \includegraphics[height=9.5cm,width=0.65\textwidth]{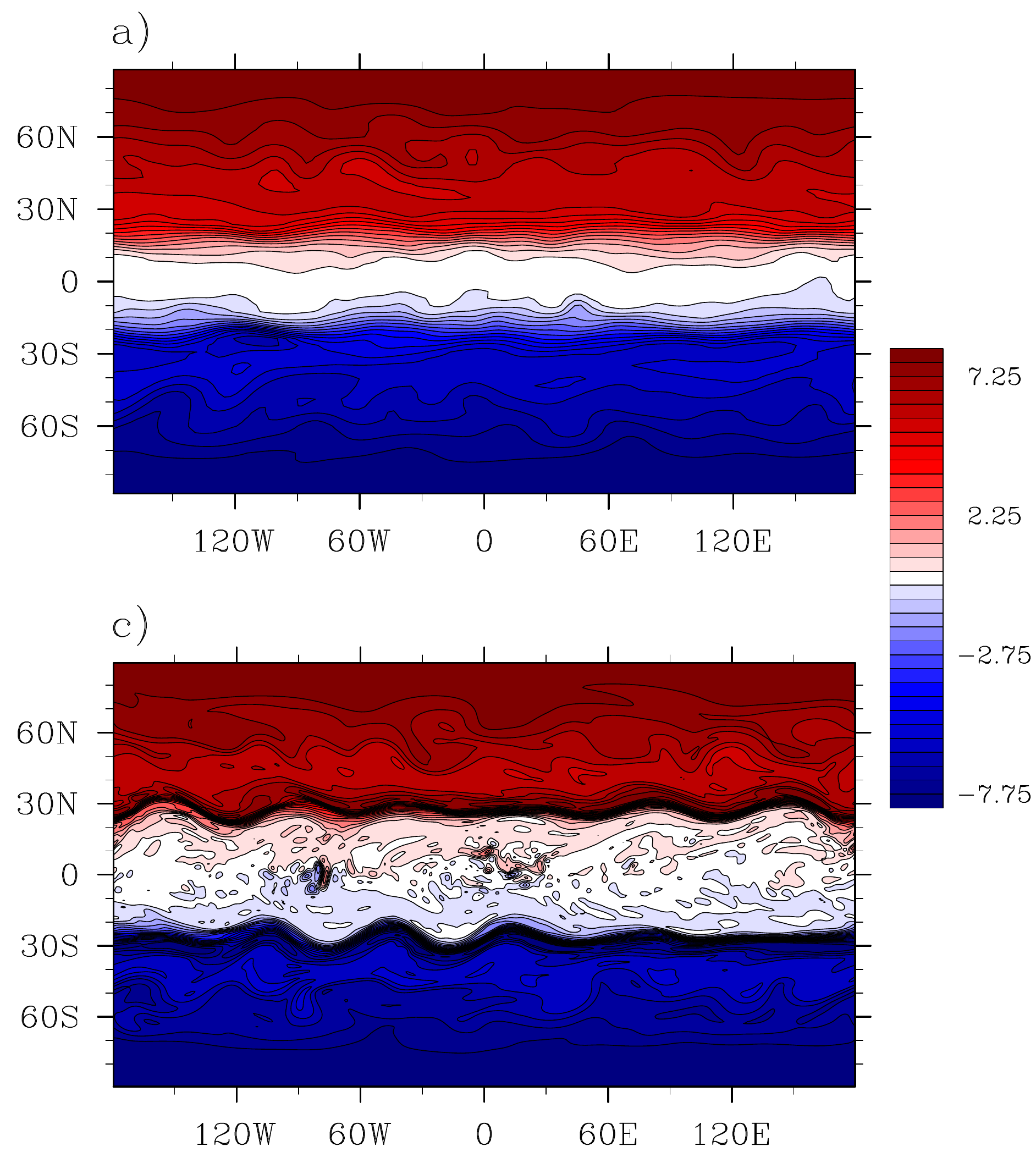}
   \includegraphics[height=9.5cm,width=0.25\textwidth]{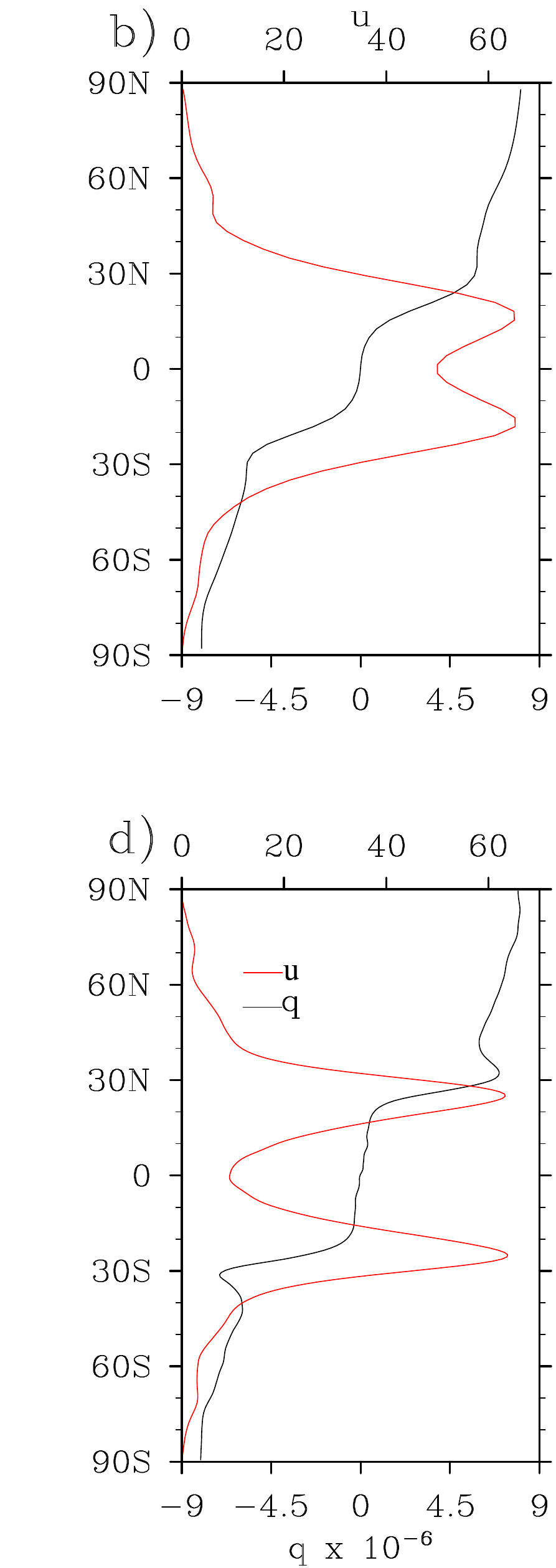}
  \end{array}$
  \caption{Potential vorticity field, $q \times
    10^{-6}$\ \,[K\,m$^2$\,kg$^{-1}$\,s$^{-1}$], in
    cylindrical-equidistant projection on the $\Theta = 330$\,K
    isentrope (approximately at $p = 200$\,hPa in the tropics) and $t
    = 5000$\,days for the RTG simulation at T42L30 ($a$) and T170L30
    ($c$) resolutions.  Notice the weaker $q$-gradients in the
    subtropics and narrower equatorial $q$-step at T42 horizontal
    resolution.  In $b$) and $d$), $\overline{q}$ (black curve) and
    zonal-mean $\overline{u}$ (red curve) for the simulations shown in
    $a$) and $c$), respectively.  At low resolution, the subtropical
    jets are more diffused and broader than at high resolution.  At
    high resolution mixing of $q$ on the equator-ward flanks of the
    jets is stronger.  As a consequence, superrotation is actually
    stronger at low resolution.  }\label{fig14}\end{figure*}

Superrotation is stronger in the lower-resolution and/or
stronger-dissipation simulation, due to diffusion of $q$. In those
simulations, $q$-contours are more diffused in the subtropics.  Hence,
the resulting jets are less sharp on their equatorward flanks and the
flow is more strongly superrotating at the equator
(cf. Figures~\ref{fig11}$a$--$d$).  The mechanism can be clearly seen
in Figure~\ref{fig14}, which shows the $q$ distributions on the
$\theta = 330$\,K isentrope from two RTG simulations (T42 and T170
resolutions) at $t = 5000$\,days.  The $q$-gradients are clearly
sharper at $\phi \approx \pm 25\degree$ and the equatorial
$q$-step\footnote{$\overline{q} = \overline{q}(\phi)$ forms a
  `staircase-like' structure, with `steps' and `jumps' \citep[see
    e.g.][and references therein]{Dritschel08}} is wider in the T170
resolution simulation -- i.e., the jets across the equator are much
less `fused' into each other. It is important to understand that this
is an entirely different mechanism than the `barotropic instability
exciting equatorial Kelvin waves' mechanism discussed in
section~\ref{sec3.2}.

In addition, in under-resolved and/or over-dissipated simulations the
meridional shear on the equatorward flanks of the subtropical jets is
reduced as the jets are `fused' into each other. Therefore it becomes
increasingly harder to satisfy the Rayleigh-Kuo instability criterion
in the time- and zonal-mean, following the acceleration stage (e.g.,
at T42 resolution this occurs at $t=400$\,days).  This results in the
mitigation of nonlinear mixing of $q$ at the equatorward flanks of the
subtropical jets. The $m\!  \approx\! 6$ eddies at $\phi \approx \pm
10\degree$--$25\degree$ (Figure~\ref{fig5}; see also
Figure~\ref{fig6}) are completely absent in the equilibrated stage in
the low-resolution/high-dissipation simulations.  Note that the Kelvin
waves can still be excited in under-resolved and over-dissipated
simulations as the $\del \overline{Q}/\del \phi$ continues to change
sign at $\phi\approx \pm 5\degree$.  The reversal of the $\del
\overline{Q}/\del \phi$ sign in this case is, however, due to the
baroclinic component. The Kelvin wave amplitude in the equilibrated
stage is, however, by as much as $\approx 30\%$ weaker than in the
converged simulations.

To summarize:
\begin{itemize}
\item{Compared to the control simulation, high (\,$\gtrsim\!$~T170)
  horizontal resolution is required for numerical convergence in the
  RTG simulation.}
\item{In low-resolution/high-dissipation RTG simulations
  superrotation is artificially enhanced as the linear diffusion of
  $q$ `fuses' the subtropical jets together on their equatorward
  flanks. This creates an artificial source of superrotation as the
  flanks of the westerly jets overlap at the equator.}
\item{Because the $q$-gradients on the equatorward flanks of the
  subtropical jets are less sharp, barotropic instability is weaker or
  absent in the under-resolved/over-dissipated RTG simulations. This
  mitigates non-linear mixing of $q$ by the barotropic $m\approx6$ eddies
  at $\phi \approx \pm 10\degree$--$25\degree$.}
\item{`Numerical diffusion' mechanism is entirely different from the
  `barotropic instability exciting Kelvin waves' mechanism for
  generating and maintaining superrotation, discussed in the preceding
  subsection.}
\end{itemize}

\subsection{Other sensitivities}\label{sec3.4}

In addition to resolution and dissipation, the strength of
superrotation in the RTG simulations is sensitive to parameters in the
equilibrium temperature distribution, such as \{$\gamma$, $b$, $\Delta
T$, $T_{\rm s}$\} in equation~(\ref{eqT}). Recall that
  the following criteria on the westerly jets must be met for
  generation of superrotation in fully resolved simulations: 1) weak
  mid-latitude baroclinic instability; and 2) shear instability within
  the deformation length scale of an equatorial Kelvin wave. If the
  specified equilibrium temperature distribution produces westerly
  jets that do not meet the above criteria, superrotation does not
  occur. For example, for $\gamma \ge 0.05$ or $b < 4$, transition to
  superrotation does not occur --- especially in
  high-resolution/low-dissipation simulations. For a larger $\gamma$
  baroclinic instability becomes stronger as the near-surface
  equator-to-pole temperature gradient is increased.  However,
  a transition is still possible if $b > 4$ for the
  same $\gamma$~($\ge 0.05$) because baroclinic instability is then
  confined to lower latitudes \citep{Williams03}. For smaller $b$ the
  subtropical jets are located too far from the equator for
  criterion~2) to be satisfied. Setting $\Delta T > 60$~K (and keeping
  all other parameters the same) also produces stronger superrotation
  because stronger meridional shear on the equatorward flanks of the
  subtropical jets more strongly satisfies the necessary criterion for
  barotropic instability than in the nominal RTG case.

The magnitude of superrotation in the RTG simulation is also somewhat
sensitive to the form of the initial perturbation.  A north-south
symmetric perturbation---such as a small amplitude localized bump at
$(\pm\!  \phi_0,\lambda_0)$ or a pure spherical harmonic mode
perturbation---produces north-south symmetric circulation and
considerably stronger superrotation than in the simulations
initialized with north-south asymmetric perturbation---such as
Gaussian white noise. In such north-south symmetric simulations, the
equatorial flow is purely zonal (i.e., does not meander) and as a
result the superrotation is stronger.

It is also worth noting that superrotation is generated in the RTG
simulation started from rest even in the absence of the Rayleigh
drag. Hence, superrotation can occur on planets with no solid surface
as long as the equator-to-pole equilibrium temperature gradient is
weakened in the region below where the superrotating flow occurs. In
summary, it appears that superrotation can be generated and maintained
under a surprisingly wide range of conditions, in general.

\section{Conclusions}\label{sec4}

In this paper, we have shown that equatorial superrotation can be
readily generated, if the equator-to-pole surface temperature gradient
is weak.  This is so even under zonally-symmetric, statically-stable,
thermal forcing in many cases.  However, in general, superrotation
does not arise as easily under zonally-symmetric forcing.
Superrotation can be maintained under this situation if
stirring from a near-equatorial barotropic instability
  is continuously provided.

When the equator-to-pole surface temperature gradient is reduced, the
role of Rossby waves produced by mid-latitude baroclinic instability
in converging easterly momentum into the equatorial region is
diminished.  Also, if a wave `source' exists at the equator (e.g. in
the form of stirring), superrotation can develop if the eddy angular
momentum fluxes out of the region are strong enough and directed up
the mean angular momentum gradient. Similar to \cite{Williams03} and
\cite{Mitchell10}, we also observe in this study that barotropic
instability generates equatorial superrotation in the
zonally-symmetric RTG setup: the instability excites equatorial Kelvin
waves, which are important for flow acceleration at the equator. In
the control simulation the necessary criterion for barotropic or
baroclinic instability is satisfied too far from the equator --
outside the deformation length scale distance of a Kelvin wave -- for
the instabilities to influence equatorial dynamics. Barotropic
instability might also be the source of Kelvin wave-like disturbances
that produce superrotation in small $a$ or $\Omega$ regimes in
\cite{potter2014}.  For a fixed phase speed, the meridional width of a
Kelvin wave increases with decreasing $a$ (and $\Omega$).  Thus, the
Rayleigh-Kuo instability criterion does not need to be satisfied as
close to the equator as in the RTG simulations with Earth's $\Omega$
and $a$, discussed in this paper. In contrast to what is suggested in
\cite{potter2014}, according to this study, it is likely that the
Kelvin wave mechanism they propose must be accompanied by a
simultaneous occurrence of barotropic (or other shear) instability, in
simulations with statically-stable, zonally-symmetric forcing.

The barotropic instability is not, however, the only mechanism
responsible for superrotation in the RTG setup.  Numerical diffusion
(both implicit and explicit) generates artificially strong
superrotation in under-resolved and/or over-dissipated RTG
simulations. In this study, it is found that to adequately resolve the
dynamics of angular momentum fluxes near the equator, relatively high
horizontal resolution is required (\,$\gtrsim\!$~T170).  In low
resolution or strongly dissipated simulations, potential vorticity is
more spread out by diffusion than in the high resolution or low
dissipation simulations. Consequently, in the under-resolved cases the
subtropical jets are less sharp on their equatorward flanks. The
flanks overlap at the equator and this provides an additional source
of superrotation. In addition, weaker potential vorticity gradients
mitigate barotropic instability and mixing by the $m\!\approx\!  6$
Rossby waves, equatorward of $\phi \approx \pm 25\degree$. This allows
the subtropical jets to further `fuse' into each other.

Given the foregoing, the relative contributions of barotropic
instability and numerical diffusion to superrotation generation can be
adequately quantified only after thoroughly assessing numerical
convergence of the simulations.  The stronger resolution requirement
for the RTG simulation, compared to the control simulation (which is
converged at T85 horizontal resolution), is due to the near-equatorial
location of the mixing region in the RTG simulations.  When the
subtropical jets (and hence the mixing zone) are located further
poleward, low resolution or high dissipation can still
induce superrotation.

\section*{Acknowledgments}
The authors thank David Dritschel, Simon Peatman, Peter Read, Ted
Shepherd and Stephen Thomson for useful discussions pertaining to this
work. The two anonymous reviewers and Ed Gerber are also thanked for
their helpful comments on the manuscript. I.P. acknowledges support by
the UK Science and Technology Facilities Council research studentship
and the hospitality of the Kavli Institute for Theoretical Physics
(KITP), Santa Barbara. J.Y-K.C.  acknowledges the hospitality of the
Isaac Newton Institute for Mathematical Sciences (Cambridge, UK) and
KITP, where some of this work was completed.

\end{document}

%% file: chocros.tex
\usepackage{amssymb}
\usepackage{upgreek}
\usepackage{bbm}
\usepackage{dsfont}
\usepackage{sansmath}
\usepackage{mathrsfs}
\usepackage{yfonts}
\usepackage{euscript}

\def\del{\partial}

\def\grad-s{\mbox{\boldmath{$\nabla$}}_{\!\! s}\,}

\usepackage{color}
\definecolor{com_red}{rgb}{.8,.2,0.1}
\definecolor{ins_green}{rgb}{.1,.5,0.1}

\newcommand{\ins}[1]{\textcolor{ins_green}{#1}}

%% file: suprot_revised_vs2.bbl
\begin{thebibliography}{99}
\bibitem[Andrews et al.(1987)]{Andrews87} Andrews DG, Holton JR, Leovy
  CB. 1987. \emph{Middle atmosphere dynamics}. No. 40. Academic press

\bibitem[Andrews and McIntyre(1978)]{Andrews78} Andrews DG, McIntyre
  ME. 1978.  Generalized Eliassen-Palm and Charney-Drazin theorems for
  waves on axisymmetric mean flows in compressible atmospheres,
  \emph{J. Atmos. Sci.}, 35, 175--185.

\bibitem[Arnold et al.(2012)]{Arnold12} Arnold NP, Tziperman E,
  Farrell B. 2012. Abrupt transition to strong superrotation driven by
  equatorial wave resonance in an idealized
  GCM. \emph{J. Atmos. Sci.}, 69,
  626--640. DOI:10.1175/JAS-D-11-0136.1

\bibitem[Asselin(1972)]{Asselin72} Asselin R. 1972. Frequency filter
  for time integrations, \emph{Mon. Wea. Rev.}, 100, 487--490. DOI:
  10.1175/1520-0493(1972)100<0487:FFFTI>2.3.CO;2

\bibitem[Chen et al.(2007)]{Chen07}Chen G, Held IM, Robinson
  WA. 2007. Sensitivity of the latitude of the surface westerlies to
  surface friction. \emph{J. Atmos. Sci.}, 64, 2899--2915. DOI:
  10.1175/JAS3995.1

\bibitem[Dritschel and McIntyre(2008)]{Dritschel08} Dritschel DG,
  McIntyre ME. 2008. Multiple jets as PV staircases: The Phillips
  effect and the resilience of eddy-transport barriers.
  \emph{J. Atmos. Sci.}, 65, 855--874. DOI: 10.1175/2007JAS2227.1

\bibitem[Hayashi(1971)]{Hayashi71} Hayashi Y. 1971. A generalized
  method of resolving disturbances into progressive and retrogressive
  waves by space Fourier and time cross-spectral analysis,
  \emph{J. Meteorol. Soc. Jpn.}, 49, 125--128.

\bibitem[Held and Hoskins(1985)]{Held85} Held IM, Hoskins BJ. 1985.
  Large-scale eddies and the general circulation of the troposphere,
  \emph{Advances in geophysics}, 28, 3--31.

\bibitem[Held and Suarez(1994)]{HeldSuarez94} Held IM, Suarez
  MJ. 1994. A proposal for the intercomparison of the dynamical cores
  of atmospheric general circulation models, \emph{Bull. A.M.S.}, 75,
  1825--1830. DOI: 10.1175/1520-0477(1994)075<1825:APFTIO>2.0.CO;2

\bibitem[Hide(1969)]{Hide69} Hide~R. 1969. Dynamics of the atmospheres
  of major planets with an appendix on the viscous boundary layer at
  the rigid boundary surface of an electrically conducting rotating
  fluid in the presence of a magnetic field, \emph{J. Atmos. Sci.},
  26, 841--853. DOI:10.1175/1520-0469(1969)026<0841:DOTAOT>2.0.CO;2

\bibitem[Holton and Lindzen(1972)]{Holton72}Holton JR, Lindzen
  RS. 1972. An updated theory for the quasi-biennial cycle of the
  tropical stratosphere. \emph{J. Atmos. Sci.}, 29, 1076--1080. DOI:
  10.1175/1520-0469(1972)029<1076:AUTFTQ>2.0.CO;2

\bibitem[Hoskins et al.(1999)]{Hoskins99} Hoskins B, Neale R, Rodwell
  M, Yang G-Y. 1999. Aspects of the large-scale tropical atmospheric
  circulation, \emph{Tellus}, 51, 33--44. DOI:
  10.1034/j.1600-0889.1999.00004.x

\bibitem[Iga and Matsuda(2004)]{Iga04} Iga S, Matsuda Y.  2005. Shear
  instability in a shallow water model with implications for the Venus
  atmosphere, \emph{J. Atmos. Sci.}, 62, 2514--2527. DOI:
  10.1175/JAS3484.1

\bibitem[Kidston and Vallis(2010)]{Kidston10} Kidston J, Vallis
  GK. 2010. Relationship between eddy-driven jet latitude and
  width. \emph{GRL}, 37. DOI:10.1029/2010GL044849

\bibitem[Laraia and Schneider(2015)]{Laraia15} Laraia AL, Schneider
  T. 2015. Superrotation in Terrestrial Atmospheres,
  \emph{J. Atmos. Sci.}, 72, 4281--4296. DOI: 10.1175/JAS-D-15-0030.1

\bibitem[Lee et al.(2005)]{Lee05}Lee C, Lewis SR, Read PL. 2005. A
  numerical model of the atmosphere of Venus. \emph{Advances in Space
    Research}, 36, 2142--2145. DOI: 10.1016/j.asr.2005.03.120

\bibitem[Matsuno(1966)]{Matsuno66} Matsuno T. 1966. Quasi-geostrophic
  motions in the equatorial area, \emph{J. Meteor. Soc. Japan}, 44,
  25--43. 

\bibitem[Mitchell and Vallis(2010)]{Mitchell10} Mitchell C, Vallis
  GK. 2010. The transition to superrotation in terrestrial
  atmospheres, \emph{J. Geophys. Res.}, 115. DOI: 10.1029/2010JE003587

\bibitem[Polichtchouk and Cho(2012)]{Polichtchouk12} Polichtchouk I,
  Cho Y-K J, 2012.  Baroclinic instability on hot extrasolar planets,
  \emph{MNRAS}, 424, 1307--1326. DOI:10.1111/j.1365-2966.2012.21312.x

\bibitem[Polvani et al.(2004)]{Polvani04}Polvani LM, Scott RK, Thomas
  SJ. 2004. Numerically converged solutions of the global primitive
  equations for testing the dynamical core of atmospheric
  GCMs. \emph{Mon. Wea. Rev.}, 132, 2539--2552. DOI: 10.1175/MWR2788.1

\bibitem[Potter et al.(2014)]{potter2014}Potter SF, Vallis GK,
  Mitchell JL. 2014. Spontaneous superrotation and the role of Kelvin
  waves in an idealized dry GCM. \emph{J. Atmos. Sci.}, 71,
  596--614. DOI: 10.1175/JAS-D-13-0150.1

\bibitem[Randel and Held(1991)]{Randel91} Randel WJ, Held IM. 1991.
  Phase speed spectra of transient eddy fluxes and critical layer
  absorption, \emph{J. Atmos. Sci.}, 48,
  688--697. DOI:10.1175/1520-0469(1991)048<0688:PSSOTE>2.0.CO;2

\bibitem[Read(1986)]{Read86} Read~PL. 1986. Super-rotation and
  diffusion of axial angular momentum: II. a review of
  quasi-axisymmetric models of planetary atmospheres,
  \emph{Quart. J. R. Met. Soc.}, 112, 253--272. DOI:
  10.1002/qj.49711247114

\bibitem[Robert(1966)]{Robert66} Robert A. 1966. The integration of a
  low order spectral form of the primitive meteorological equations,
  \emph{J. Met. Soc. Japan}, 44, 237--245.

\bibitem[Sakai(1989)]{Sakai89} Sakai S. 1989. Rossby-Kelvin
  instability: a new type of ageostrophic instability caused by a
  resonance between Rossby waves and gravity waves. \emph{JFM}, 202,
  149--176. DOI: 10.1017/S0022112089001138

\bibitem[Saravanan(1993)]{Saravanan93}Saravanan R. 1993. Equatorial
  superrotation and maintenance of the general circulation in
  two-level models. \emph{J. Atmos. Sci.}, 50,
  1211--1227. DOI:10.1175/1520-0469(1993)050<1211:ESAMOT>2.0.CO;2

\bibitem[Schneider and Liu(2009)]{Schneider09}Schneider T, Liu
  J. 2009. Formation of Jets and Equatorial Superrotation on
  Jupiter. \emph{J. Atmos. Sci.}, 66, 579--601. DOI:
  10.1175/2008JAS2798.1

\bibitem[Scott et al.(2003)]{Scott03} Scott RK, Rivier L, Loft R,
  Polvani LM. 2003. \emph{`BOB: model description and users guide'},
  NCAR Technical Note, 32pp, NCAR: Boulder, Colorado, USA.

\bibitem[Suarez and Duffy(1992)]{Suarez92} Suarez MJ, Duffy DG. 1992:
  Terrestrial superrotation: A bifurcation of the general
  circulation. \emph{J. Atmos. Sci.}, 49, 1541--1554. DOI:
  10.1175/1520-0469(1992)049<1541:TSABOT>2.0.CO;2

\bibitem[Tziperman and Farrell(2009)]{Tziperman09} Tziperman E,
  Farrell B. 2009. Pliocene equatorial temperature: Lessons from
  atmospheric superrotation. \emph{Paleoceanography},
  24. DOI:10.1029/2008PA001652

\bibitem[Wang and Mitchell(2014)]{Wang14} Wang P, Mitchell
  JL. 2014. Planetary ageostrophic instability leads to
  superrotation. \emph{GRL}, 41, 4118-4126. DOI: 10.1002/2014GL060345

\bibitem[Wheeler and Kiladis(1999)]{wheeler1999} Wheeler M, Kiladis
  GN. 1999. Convectively coupled equatorial waves: Analysis of clouds
  and temperature in the wavenumber-frequency domain,
  \emph{J. Atmos. Sci.}, 56,
  374--399. DOI:10.1175/1520-0469(1999)056<0374:CCEWAO>2.0.CO;2

\bibitem[Williams(2006)]{Williams06} Williams~GP. 2006. Equatorial
  Superrotation and Barotropic Instability: Static Stability Variants,
  \emph{J. Atmos. Sci.}, 63, 1548--1557. DOI:10.1175/JAS3711.1

\bibitem[Williams(2003)]{Williams03} Williams~GP. 2003. Barotropic
  instability and equatorial superrotation, \emph{J. Atmos. Sci.}, 60,
  2136--2152. DOI:10.1175/1520-0469(2003)060<2136:BIAES>2.0.CO;2

\bibitem[Williams(1988)]{Williams88} Williams~GP. 1988. The dynamical
  range of global circulations--I. \emph{Climate Dyn.}, 2,
  205--260. DOI:10.1007/BF01371320

\bibitem[Yamamoto and Takanashi(2003)]{Yamamoto03} Yamamoto~M,
  Takahashi M. 2003. The fully developed superrotation simulated by a
  general circulation model of a Venus-like
  atmosphere. \emph{J. Atmos. Sci.}, 60,
  561--574.DOI:10.1175/1520-0469(2003)060<0561:TFDSSB>2.0.CO;2
\end{thebibliography}
